\documentclass[modern]{aastex63}
\usepackage{amssymb, amsmath}
\usepackage{color}
\usepackage{graphicx}
\usepackage{morefloats}
\usepackage{hyperref}
\NewPageAfterKeywords
\begin{document}

\title{Testing Evolutionary Models with Red Supergiant and Wolf-Rayet Populations}

\author{Philip Massey}
\affiliation{Lowell Observatory, 1400 W Mars Hill Road, Flagstaff, AZ 86001, USA}
\affiliation{Department of Astronomy and Planetary Science, Northern Arizona University, Flagstaff, AZ, 86011-6010, USA}
\email{massey@lowell.edu,kneugent@lowell.edu,tzdw@uw.edu,j.eldridge@auckland.ac.nz,\\e.r.stanway@warwick.ac.uk,emsque@uw.edu}

\author{Kathryn F. Neugent}
\affiliation{Lowell Observatory, 1400 W Mars Hill Road, Flagstaff, AZ 86001, USA}
\affiliation{Department of Astronomy, University of Washington, Seattle, WA, 98195, USA}

\author{Trevor Z. Dorn-Wallenstein}
\affiliation{Department of Astronomy, University of Washington, Seattle, WA, 98195, USA}

\author{J. J. Eldridge}
\affiliation{Department of Physics, University of Auckland, Private Bag 92019, Auckland, New Zealand}

\author{E. R. Stanway}
\affiliation{Department of Physics, University of Warwick, Gibbet Hill Road, Coventry, CV4 7AL, UK}

\author{Emily M. Levesque}
\affiliation{Department of Astronomy, University of Washington, Seattle, WA, 98195, USA}

\begin{abstract}

Despite the many successes that modern massive star evolutionary theory has enjoyed, reproducing the apparent trend in the relative number of red supergiants (RSGs) and Wolf-Rayet (WR) stars has remained elusive.  
Previous estimates show the RSG/WR ratio decreasing strongly with increasing metallicity.  However, the evolutionary models have always predicted a relatively flat distribution for the RSG/WR ratio.  In this paper we reexamine this issue, drawing on recent surveys for RSGs and WRs in the Magellanic Clouds, M31, and M33.  The RSG surveys have used Gaia astrometry to eliminate foreground contamination, and have separated RSGs from asymptotic giant branch stars using near-infrared colors.  The surveys for WRs have utilized interference filter imaging, photometry, and image subtraction techniques to identify candidates, which have then been confirmed spectroscopically.  After carefully matching the observational criteria to the models, we now find good agreement in both the single-star Geneva and binary BPASS models with the new observations.  The agreement is better when we shift the RSG effective temperatures derived from $J-Ks$ photometry downwards by 200~K in order to agree with the Levesque TiO effective temperature scale.  In an appendix we also present a source list of RSGs for the SMC which includes effective temperatures and luminosities derived from near-infrared 2MASS photometry,
in the same manner as used for the other galaxies.

\end{abstract}

\section{Introduction}

We often refer to massive stars as ``cosmic engines" as they are responsible for manufacturing many of the heavy elements in the Universe \citep{JJohnson}.  In addition, their strong stellar winds, and eventual disruption as core-collapse supernovae, provide most of the mechanical energy input to the interstellar medium, triggering new generations of star formation (see, e.g., \citealt{Dave}).  Further, their UV radiation creates the H\,{\sc ii} regions that populate late-type galaxies. Massive stars are also the source of the most energetic phenomenon yet found, emitting gamma-ray bursts (GRBs) as they collapse into black holes \citep{1993ApJ...405..273W,  2002AJ....123.1111B}.   Mergers of their stellar remnants (black holes and neutron stars) disturb the curvature of spacetime, propagating as gravitational waves.  Understanding massive star evolution, and getting the evolutionary models {\it right} (and knowing their limitations) is  important not only for understanding the evolution of massive stars {\it per se}, but also for interpreting the light from distant galaxies (e.g., \citealt{2017IAUS..329..305S}), and a host of other astrophysically interesting phenomena.

Massive star evolution is not a solved problem. The physics of massive stars pose many challenges to modeling their evolution;  chief among these is mass loss. The first stellar UV spectroscopic observations showed that mass loss is ubiquitous among massive OB stars \citep{1967ApJ...147.1017M,1977ApJS...33..269S}.  The incorporation of mass loss in early stellar models showed the dramatic effect this had on their evolution (see, e.g., \citealt{1974A&A....34..355C,1977A&A....61..251D,1978A&AS...34..363D,1979A&A....74...62C,1980A&A....87...68L,1980SSRv...26..113D,1982ApJ...256..247B,1982ApJS...49..447B}), broadening the main-sequence and causing stars to evolve at higher luminosities than they would without mass loss.   It also provides a mechanism to produce Wolf-Rayet (WR) stars through single-star evolution\citep{1975MSRSL...9..193C,1979A&A....74...62C,1986ARA&A..24..329C}. Aside from mass loss, rotation is another important ingredient in the models.  Although we often think of stellar evolution as being determined by the initial mass and chemical composition (i.e., the Russell-Vogt theorem), the initial angular momentum has also been shown to be a third important parameter due primarily to effects of rotationally induced mixing, but its inclusion in evolution is complicated. (See, e.g., \citealt{2000ARA&A..38..143M} and \citealt{2009pfer.book.....M}.)   Finally, mass transfer and tidal interaction due to a binary companion will also have profound effects on the subsequent evolution of massive stars, as reviewed by \citet{2012ARA&A..50..107L} and  \citet{2017PASA...34....1D}.  Inclusion of interacting binaries in massive star evolution models strongly affects the resulting stellar populations; see, e.g., \citet{2020arXiv200511883E}.

Despite these complications, the current generation of evolutionary models have enjoyed many successes in matching the observed distributions and properties of massive stars.  Single-star evolutionary models have been shown to match the luminosity distribution of the very short-lived yellow supergiant phase in the LMC and in M33 \citep{NeugentLMC,DroutM33}.  Both single-star and binary evolution models match the relative number of WN- and WC-type Wolf-Rayet (WR) stars  in nearby galaxies \citep{NeugentM31,BPASS2,2018ApJ...867..125D,2020MNRAS.497.2201S}, as well as the shape of the luminosity function of RSGs in M31 \citep{UKIRT}.  It should be emphasized that such tests involving the numbers of evolved massive stars is a very exacting one, as it turns a
``magnifying lens" on stellar evolution calculations \citep{Kipp,MeynetMaeder05}.  

Possibly the most vexing failure has been the inability of these models to correctly predict the relative number of RSGs and WRs. 
In the standard single-star evolutionary models (\citealt{Sylvia, BPASS2}; P. Eggenberger et al.\ 2021, in prep.) there is some mass limit 
($\sim$30$M_\odot$) below which massive stars evolve to the RSG phase, and above which they evolve to the WR phase, possibly with a narrow band of masses where a RSG may become a WR. (For a somewhat different view, see \citealt{2018ApJS..237...13L}.)
This upper mass for evolution to the RSG phase is set by the Eddington limit: stars that are more massive (and hence luminous) than some value cannot cross the HR diagram to cool, RSG-like temperatures, as opacities in their outer layers increase as the star evolves to cooler temperatures. At some point outwards radiation pressure approaches the pull of gravity, and the star
belches out its cooler layers.  This phase is often identified as the {\it luminous blue variable} (LBV) phase (see, e.g., \citealt{1988ApJ...324..279L}).  This process is revealed in the HR diagram as the Humphreys-Davidson limit, the band of upper luminosity first  characterized by \citet{1979ApJ...232..409H}.  This extends from the highest luminosities and effective temperatures diagonally to $\log L/L_\odot\sim$5.6 and $T_{\rm eff}\sim$15,000~K, and then horizontally (see Figure~22.1 in \citealt{LamersLevesque}).  In this classical, single-star picture, stars with masses above the RSG cutoff lose enough mass through stellar winds and (for the most massive stars) the LBV phase that the bare core of the star is exposed, and a WR star is born.  In this picture, RSGs are the He-burning descendants of 9-30$M_\odot$ stars and have ages of 8-35~Myr \citep{Sylvia}, while WRs come from higher mass stars and have ages of 2-6~Myr \citep{ 2012A&A...542A..29G}.  Even when binary evolution is included, WR stars have ages $<$10~Myr \citep{2008MNRAS.384.1109E}, still shorter than most RSGs.

 \citet{Maeder80} argued that the {\it observed} ratio of RSGs to WRs showed a very strong gradient with metallicity, increasing by two orders of magnitude within the the Galaxy as one went from galactocentric distances of 7-9~kpc to 11-13~kpc, with a presumed decrease in metallicity.   Furthermore this trend seemed to be consistent with extension to the lower metallicities of the Large and Small Magellanic Clouds (LMC, SMC).  Although the scaling of mass-loss rates with metallicity was unknown at the time, it was understood that these stellar winds were driven by radiation pressure in highly ionized metal lines; thus, mass-loss rates should be lower in stars of lower metallicity.  This would mean that stellar winds would be more effective at producing WRs at higher metallicities.  The Humphreys-Davidson limit was unrecognized at the time, and it was thought that RSGs were an earlier phase in the lives of most WRs, with stars with masses of 15-60$M_\odot$ passing through both phases.  Thus \citet{Maeder80} proposed that the observed metallicity dependence in the RSG/WR ratio was due to the effect that different mass-loss rates would have on the relative lifetimes of the RSG and WR phases.  However, even with the modern understanding that RSGs and WRs came from different mass ranges, the influence of metallicity on mass-loss rates still provided a reasonable explanation, as one might expect that the luminosity (mass) limit for RSGs would be higher in lower metallicity environments, and hence the RSG/WR ratio would vary with metallicity.
 
Later, when stellar evolutionary models with scaling of mass-loss rates by metallicity were available, \citet{MasseyARAA} found that (surprisingly) these early stellar evolution models really did not predict much of an expected change in the RSG/WR ratio at all.  In Figure~\ref{fig:araa} we show how poorly the data and models compared at the time.  Still, those models included neither rotation nor binarity, and the presumption was that as the models improved, so would the agreement, but that has not been the case.    \citet{2008MNRAS.384.1109E} found that even by including binaries, that the expected relationship in the RSG/WR ratio with metallicity remained relatively flat compared to the observed distribution (see their Figure~6).  Similar disagreement can be seen in Figure~16 of \citet{BPASS2}, which used an improved version of the binary models as well as a better estimate of the observed RSG/WR ratio provided by the first author of the present paper.   Inclusion of rotation in the single star models, and accounting for the effect of binarity in producing WRs, continued to show flatter distributions in the expected RSG/WR ratios with metallicity than  the observations \citep{Fred20,2020MNRAS.497.2201S}.   

However, much has improved observationally in the decades since the \citet{Maeder80} and \citet{MasseyARAA} studies.  New surveys for both RSGs and WRs have been carried out amongst the Local Group galaxies, with the result that we now have samples whose completeness are well understood (Section~\ref{Sec-sample}).  Thus, we use this opportunity to present significantly improved values for the RSG/WR ratios as a function of metallicity (Section~\ref{Sec-Ratios}).  We then go on to compare these ratios to two of the most used sets of evolutionary models, carefully describing our procedures so that others may follow in our footsteps with other evolutionary models, available either now or those of the future (Section~\ref{Sec-Models}).  We summarize our results and discuss the implications in Section~\ref{Sec-Sum}.

 \section{Sample Selection}
\label{Sec-sample}

In order to compare the observed RSG/WR ratios with those predicted by the evolutionary models, we must make sure that the range of physical properties in our sample is well enough understood to correctly identify the corresponding phases in the tracks.  The models give us the effective temperatures, luminosities, and surface compositions as a function of time, and in order to make valid comparisons, we must match these criteria to the physical properties of our observational sample.   As we show below, this is relatively straightforward for the RSGs, as our samples have well-defined temperature and luminosity limits.  However, for WR stars this is more complicated, as WRs are recognized observationally on the basis of emission lines formed in their stellar winds, and not simply on the basis of effective temperatures or luminosities.  Fortunately, the surface chemical abundances, combined with temperature, provide the key for recognizing this phase in the evolutionary models.  

Surveys for RSGs and WRs are now complete for the SMC, LMC, M31, and M33. In this section we will describe the source lists, and how we ascertain that we are identifying the RSG and WR phases in the models in the same way we have identified these stars observationally. 

\subsection{RSGs}
\label{Sec-physical}
The RSG populations of these four galaxies have all been recently identified 
using $K_s$ vs.\ $J-K_s$ color-magnitude diagrams (CMDs).  Early work by e.g., \cite{MasseyOlsen,NeugentLMC,DroutM33} used either optical or near-infrared (NIR) CMDs to select RSG candidates in nearby galaxies.
Radial velocity measurements were then used to remove foreground stars.  However, a remaining concern was the contamination of the fainter stars in their sample by LMC asymptotic giant branch (AGB) stars. Contamination by AGBs could 
only be avoided by using a high luminosity cut-off (typically $\log L/L_\odot>4.9$), as first recommended by \citet{Brunish86}.    Subsequently, the removal of foreground stars was greatly simplified by the availability of Gaia proper motions and parallaxes, as used by \cite{ErinLBV} to obtain a clean LMC RSG sample, and the AGB contamination problem was solved by \cite{Yang2019}, who demonstrated that AGBs and RSGs in the SMC could be distinguished on the basis of their $J-K_s$ colors, in accord with the earlier theoretical and observational work of \cite{2006AA...448...77C,2006A&A...452..195C} and \cite{2011AJ....142..103B}.   Using these techniques, luminosity-limited samples of RSGs have now been identified over the whole of the LMC \citep{LMCBins}, and  the two spirals M31 and M33 \citep{M3133RSGs}, and we adopt these source lists here.   

Although \cite{Yang2019} identified a sample of RSGs in SMC, we have had to redo the selection here, for the following two reasons.  The first concerns the lower limit used in $J-K_s$.   \cite{Yang2019} used a lower limit  cutoff which paralleled the sloped upper limit cutoff used to separate RSG and AGBs.  As \citet{LMCBins} argues, there is no physical basis for a sloped lower limit, and using one has the potential to overlook RSGs with hot companions at the bright end.  
The second problem is that  \cite{Yang2019}  did not convert the photometry to physical properties, which we require for our analysis here.  (The recent identification of RSGs in M31 and M33 by \citealt{MingM31M33} also suffered from these issues, and so we adopt the \citealt{M3133RSGs} list for those galaxies.)  Thus in the appendix of this paper we present our own source list of SMC RSGs.

Recognizing the RSG phase in the evolutionary tracks is relatively straightforward as long as we have a clear understanding of the luminosity and temperature criteria used in our observations. As emphasized in \citet{M3133RSGs}, it makes little sense to talk about ``the number of RSGs" in a galaxy without specifying the luminosity limit of the list.  In addition, we must also be careful to match the temperature regime for what we are calling red supergiants with the models.  The $J-K_s$ cut-offs that have been used for the source lists differ slightly from galaxy to galaxy, due to the change in the typical RSG temperature shifting to warmer temperatures at lower metallicities.  \citet{Elias1985} first noted the shift towards earlier spectral types among the RSGs as one went from Galactic to LMC to SMC, and correctly attributed this to the changes in the Hayashi limit \citep{HayashiHoshi} with metallicity. (See, e.g., Section 3.2 in \citealt{LevesqueRSGs}.)  

In Table~\ref{tab:RSGcriteria} we list the corresponding temperature ranges as a function of luminosity used in the four source lists. There does remain one uncertainty in relating this to the effective temperatures given by the evolutionary models.  The conversion of de-reddened $J-K_s$ colors to effective temperatures comes about through the extension of the one-dimensional (1D) {\sc marcs} atmosphere models \citep{Marcs75,Marcs08,Marcs92} to the low surface gravities appropriate for supergiants ($\log g {\rm [cgs]} \sim 0$), as described in  \citet{Levesque2005}.  However, as noted by \citet{Levesque2005} and \citet{Levesque2006}, and subsequent studies (see, e.g., \citealt{Davies2013,LMCBins}), there is a systematic difference between the effective temperatures derived from fitting the TiO bands using the MARCS models, and fitting de-reddened broad-band colors, in the sense that the spectroscopic temperatures are cooler. (The size of this discrepancy using $(V-K)_0$ is 50-150~K according to \citealt{Levesque2005, Levesque2006}.) This problem has been attributed to the static 1D nature of the models \citep{Levesque2005}.  \citet{Davies2013} argues that that the photometric (warmer) temperatures are more reliable as the continuum radiation is formed at deeper layers. However, the TiO spectroscopic temperatures are in 
better agreement with the evolutionary model temperatures (see, e.g., Figure 7 in \citealt{Davies2013}), although this admittedly dependent upon the choice for $\alpha$, the mixing-length parameter \citep{1987A&A...182..243M,2018ApJ...853...79C}. The TiO temperature determinations are also immune to uncertainties in the reddening corrections, which is an important consideration given the presence of circumstellar material around RSGs \citep{Smoke}.  A recent
study by \citet{Ironman} reexamined the RSG temperature scale using a novel scaling of well-established red giant temperatures and the line-depth ratio method applied to Fe\,{\sc i} lines, finding excellent agreement with the TiO method
and the evolutionary models.  (Note that their study is independent of the {\sc marcs} models.) In an appendix,  \citet{LMCBins} looked specifically at the differences between the temperatures derived from their TiO band fitting and their $(J-K)_0$ temperatures. The offset is about 200~K (see their Figure 7), again in the sense that the TiO band temperatures
are smaller (cooler) than the ones derived from photometry.  Thus in our comparison below, we will use both the nominal (warmer) values in identifying the RSG phase in the evolutionary models, and also a corrected version, found by subtracting 200~K from the photometric $(J-K)_0$ temperatures.  Based on the fact that the latter is in better agreement with where stars fall on the evolutionary tracks, these would appear to be the more appropriate ones to use in identifying the RSG phase in the models.

We do note that despite the uncertainty in the effective temperatures derived from the $J-K_s$ photometry, this has a negligible effect on the derived luminosities. As described by \citet{LMCBins}, the effect is on the order of 0.05~dex, comparable to the uncertainty in the values due to the uncertainties in the reddening corrections.

\subsection{WRs}
The history of WR surveys in Local Group galaxies has been recently reviewed by \citet{2019Galax...7...74N}.  Early photographic objective prism and interference-filter surveys were woefully incomplete, particularly for WN-type WRs, which have weaker lines (see, e.g., discussion in \citealt{MJ98}).  When CCDs first came along, the sensitivity became good enough to achieve completeness at 1~Mpc distances (i.e., within the Local Group) on 4-meter class telescopes, but the field of views were tiny compared to the angular extent of most of the nearby galaxies \citep{1985ApJ...291..685A,MAC86,MJ98}, particularly M31 and M33.  These issues were finally solved when the Mosaic CCD camera was deployed on the Kitt Peak Mayall telescope.  Such studies also benefited from the development of image subtraction techniques in order to better identify WR candidates.  Multi-object spectroscopy also facilitated spectroscopic confirmations.  Surveys for WRs in M33 and M31 were carried out by \citet{NeugentM33} and \citet{NeugentM31}, respectively, with completeness limits of 95\% within the survey areas.  (The 5\% loss came from gaps between the chips in the CCDs.)  The sensitivity of these surveys would not have been good enough to detect the WN3/O3 types, which are visually fainter \citep{MasseyMCWRI,NeugentWN3O3s,FinalCensus}, but should be complete for all other types.
The Magellanic Clouds have presented special challenges for WR surveys because of their large angular extents.    
Although a 1-meter telescope has plenty of sensitivity to detect even the weakest lined WRs at the distances of the 
Magellanic Clouds (0.05-0.06~Mpc), each of the Clouds covers a large swath of sky, making complete surveys a daunting task, particularly for the LMC.  However, after a multi-year effort, both Magellanic Clouds have now been completely surveyed and candidates examined spectroscopically \citep{MasseyMCWRI,MasseyMCWRII,MasseyMCWRIII,FinalCensus}.  

How do we recognize the WR phase in evolutionary tracks?  As mentioned above, the ``WR phenomenon" is atmospheric, in the sense that very broad emission lines of He and N (the products of CNO hydrogen-burning), or C and O (the products of He fusion),  are seen in the spectrum.  The associated effective temperatures are high \citep{1987ApJS...63..947H,1988ApJ...327..822H,1989ApJ...347..392H} and the surface hydrogen abundance low \citep{1983ApJ...268..228C}.  In his review, \citet{2007ARA&A..45..177C} cites numbers of $T_{\rm eff}$=30,000-100,000~K and typical mass fractions $X_H$ less than 0.15.  Traditionally, theorists have used the relatively simple criteria that a model star with a high effective temperature $>$10,000-28,000~K and a hydrogen mass fraction $X_H$$<$$0.3-0.4$ can be assumed to be in the WR phase \citep{MeynetMaeder03,CyrilPops,BPASS2,2018ApJS..237...13L}. 
Changing these conditions slightly (such as adopting $X_H$$<$$0.4$ vs.\ 0.3, or the lower limit to the effective temperature) do not change the statistics significantly. 
\citet{BPASS2} includes the additional criterion that $\log L/L_\odot>4.9$ as in their models binary stripping can produce WR-like surface abundances but such stars lack the stellar winds that would cause them to be classified observationally as WRs.  (\citealt{2020MNRAS.497.2201S} explores the implications of including a metallicity-dependent luminosity criteria, following \citealt{2020A&A...634A..79S}.)\footnote{For evolutionary models including intermediate- and low-mass stars, and extending to a post-AGB phase, such a luminosity criterion also is useful for distinguishing Population I WRs from the WR-like central stars of planetary nebulae (see, e.g., \citealt{2020ApJ...898...85M} and references therein).}

However, there are still three types of ``slash" WR stars that we must consider whether or not to count.  For instance, what about the late-type, high-luminosity, hydrogen-rich WN stars?  These are stars that have classifications like O2-3/WN5-6, and are found in the cores of very young, rich clusters, such as the LMC's R136, the Milky Way's NGC~3603, and M33's NGC~604.  The discovery of these in NGC~604 caused \citet{CM81} to speculate that these ``superluminous" WRs (including the similar ones in R136 and NGC~3603) were some sort of unstable massive star and not true classical WRs. The possibility that they were simply unresolved multiple stars remained until {\it HST} demonstrated otherwise (see, e.g., \citealt{MH98}).  They are now believed to still be in a hydrogen-burning phase \citep{deKoterR136,MH98}, showing WR emission lines simply because they are so close to their Eddington limits that their stellar winds are optically thick \citep{1998MNRAS.296..622C,2011A&A...535A..56G,1998MNRAS.296..622C,2019A&A...627A.151S,2020MNRAS.499..873S}.   (Details of their classifications can be found in \citealt{2011MNRAS.416.1311C}.)   Quantitative analysis suggests hydrogen depletion but with mass fractions as high as $X_H=0.6-0.7$ \citep{1998MNRAS.296..622C, 2010MNRAS.408..731C}.  Thus, although such stars are typically included in WR catalogs \citep{NeugentM33,FinalCensus}, they are not what we consider to be ``real" WR stars, and would not be identified as WRs in the models.  Therefore, we will exclude these from our counts. We do note that there are very few of these stars, i.e., a few percent (2\% in M33, 6\% in the LMC), as described below in Section~\ref{Sec-WRnumbers}.

A second type of ``slash" WRs are the Ofpe/WN9 stars \citep{1989PASP..101..520B}, stars whose properties are intermediate between O-type supergiants and low excitation WNs, and which are sometimes alternatively classified as WN9-11 \citep{1995A&A...293..172C,1997A&A...320..500C}.  These stars were discovered to have the surprising tendency to undergo dramatic changes in their spectral appearance as they suddenly brightened, and many now consider the Ofpe/WN9 class to be the quiescent state of (some) LBVs \citep{2008ApJ...683L..33W,2017AJ....154...15W}.   Thus, these stars underscore one potential issue:  do we count the handful of stars that were Ofpe/WN9 stars 30 years ago but are now LBVs?  In general, evolutionary models are not refined enough to clearly recognize an LBV stage, as again the physics of what happens in the atmosphere dominates.  We will include such stars in our counts, although we recognize that a few may be missed if they are in an LBV state that may not have been noticed.  Fortunately, the percentage of such objects is also quite low, $\sim$3\% in the LMC.

Finally, there is the question of a third type of ``slash" WR, the newly discovered WN3/O3s.  Nine of these stars were recently discovered in the LMC as part of a deep search for WRs \citep{FinalCensus}. (A tenth star, LMCe055-1,  shows a spectrum that would be better described as WN4/O4; see \citealt{MasseyMCWRIII}.)   These stars show absorption lines typical of an early O-type stars but the emission of a high excitation WN star.  Modeling shows that the spectra come from a single object with temperatures and bolometric luminosities similar to WN3 stars, but with mass-loss rates more similar to that of O-type stars \citep{NeugentWN3O3s}.  They are likely a short-lived ``missing link" stage in the evolution of O stars to the WN stage, although alternatively they could be the result of binary stripping (see discussion in \citealt{NeugentWN3O3s}).   These stars have $X_H$ of about 0.2 \citep{NeugentWN3O3s}, and high effective temperatures, and thus they should likely count as WRs.  However, we note that we may be missing some of these still observationally in M31 and M33, as these stars are visually fainter ($M_V=-3$) and have weaker lines than do most WRs.  (A deeper survey is in progress, and has identified 9 potential candidates in M33, one of which has so far been confirmed spectroscopically).  Still, the relative number is low in the LMC; only 10 of the 154 (6\%) of the known WRs fall into this camp. No WN3/O3s are found in the SMC, likely because of its small number (12) of WRs. None are known in the Milky Way, and so we suspect that M31 (with a similar metallicity) does not contain any, and thus any observational incompleteness is likely to be small.  We include these in our counts for the LMC and M33.

\subsection{Spatial Coverage}

In selecting our samples of RSGs and WRs we must also make sure that the samples cover the same spatial extent.  For the Magellanic Clouds, this was straightforward, 
as we had the luxury of designing the RSG studies to match the WR survey regions.  Figs.~\ref{fig:smc} and \ref{fig:lmc} show the RSG survey region (yellow circle) and the WR survey fields (green boxes).  

In Figs.~\ref{fig:m31} and \ref{fig:m33}  we show the survey regions for M31 and M33.  The areas surveyed for the RSGs are outlined in yellow, while the surveyed regions for WRs are outlined in green.   The WR survey regions are actually about 5\% smaller than what are shown, in that there are gaps between the 8 CCDs that made up the Mosaic detectors (see, e.g., \citealt{NeugentM33,NeugentM31}) .
Since we are also interested in investigating how the RSG/WR ratio changes
with metallicity, and since we expect metallicity to be correlated with galactrocentric distance $\rho$ within the planes of these galaxies, it makes sense to restrict ourselves to the survey regions that are complete to a particular $\rho$.  In M31 (Figure~\ref{fig:m31}) the red oval denotes $\rho=0.75$, where $\rho$ has been normalized to an isophotal radius corresponding $\mu_B=25$ mag arcsec$^2$, denoted R25\footnote{For M31, we assume an R25 radius of 95\farcm3, an inclination 77\fdg0, and a position angle of the major axis of 35\fdg0 \citep{RC3}, following \citet{NeugentM31}.  At a distance of 760~kpc \citep{vandenbergh2000}, a value  of $\rho=1.0$ corresponds to galactocentric distance of 21.07~kpc. We note that in previous papers we have often incorrectly referred to R25 as the ``Holmberg radius," which actually corresponds to a surface brightness of $\mu_B=26.5$ mag arcsec$^2$ \citep{Holmberg58}. In fact, have always used the $\mu_{\rm 25}$ value from \citet{RC3} as our normalization factor, as we do here, and PM apologizes for any confusion caused by the misuse of the nomenclature.}.  We see that $\rho=0.75$ falls just within the green Mosaic fields that denote the WR survey.   For M33 (Figure~\ref{fig:m33}) our survey for WRs extends all the way out to $\rho=1.0$ thanks to the smaller angular size of the galaxy and the more favorable inclination\footnote{For M33, we assume an R25 radius of 30\farcm8 \citep{RC3}, an inclination of 56\fdg0 deg, and a position angle of the major axis of 23\fdg0 \citep{1989AJ.....97...97Z}, following \citet{NeugentM33}. At a distance of 830~kpc \citep{vandenbergh2000}, a value of $\rho=1.0$ corresponds to a galactocentric distance of 7.44~kpc.}.

In studies of individual stars in other galaxies, crowding is always a concern, even within the Local Group.  The presence of a close line-of-sight OB star of comparable luminosity to the WR star would lessen our detection limit.  This issue was examined carefully for M33 by \citet{MJ98} and \citet{NeugentM33}, who concluded that in M33 this was an issue only in the crowded cores of NGC~604 and NGC~595, and that these regions were surveyed for WRs using {\it HST} by \citet{1993AJ....105.1400D}.  The same holds for the LMC, where the crowded R136 region of 30~Dor required {\it HST} to identify the individual O2-3/WN5-6 \citep{MH98}. The faintest WRs have $M_V$$\sim$$ -3$ \citep{2007ARA&A..45..177C}, and thus any line-of-sight companion that would mask the detection of the WR would have to be very luminous, likely $M_V<-5$, i.e., a supergiant of some type. Such stars are extremely rare.  As for the RSGs, \citet{M31M33RSGBins} showed that only a few percent of RSGs would be coincident with even B dwarfs, and such instance would not, in any event, mask the detection of the RSG in the NIR.
 
For M31 there is one additional concern. \citet{M3133RSGs} found that there were far fewer matches between their NIR-selected sample of RSGs and the optical Local Group Galaxy Survey (LGGS, \citealt{LGGSI,BigTable}) than in M33.  This could be due to extinction within the plane of M31, causing fewer optical detections on the far side of M31's disk.  \citet{M3133RSGs} argues against this interpretation but it remains a concern.  We therefore follow \citet{M31M33RSGBins} in considering just the subsample of fields where there are a high percentage ($>$50\%) of matches.  As we will see below, the results in terms of the RSG/WR ratio are essentially indistinguishable full sample and this subsample.

\clearpage

\subsection{Star Formation Concerns}
\label{Sec-IMF}

The number of RSGs and WRs in a galaxy today are a reflection not only of stellar evolutionary processes but also of the initial mass function (IMF) and  star formation histories (SFHs).    The evolutionary models give us only the lifetimes of stars in a particular evolutionary stage.  In order to transform this information to a number of stars, we must adopt an IMF, and make assumptions about the star formation histories of these galaxies.   

Studies of the resolved stellar content of coeval OB associations have generally shown that the high-mass end of the IMF is  universal, with a \citet{1955ApJ...121..161S} power-law slope of $\Gamma=-1.35$, over a range of metallicities and star-formation rates  (\citealt{MasseyGilmore,2011ASPC..440...29M} and references therein).   The only regions where this may not be true are the extreme starbursts regions such as 30~Dor and NGC~3603, where some studies have argued that the IMF is top-heavy (\citealt{2012MNRAS.422.2246M,2012A&A...547A..23B,2020ApJ...904...43H} and references therein). Others studies of these same regions have found either normal IMF slopes, or have argued that the data are too incomplete to draw any conclusions (see, e.g., \citealt{MH98} for 30~Dor and \citealt{2008AJ....135..878M} for NGC~3603).

We must also understand the SFHs of these galaxies over the past 30~Myr.   The typical ages of RSGs range from 8~Myr to 35~Myr \citep{Sylvia}, while WRs have ages of 2~Myr to 10~Myr \citep{2008MNRAS.384.1109E,2012A&A...542A..29G}.  To compare the number of RSGs to WRs we generally assume that the star formation rate (SFR)  within a galaxy has been steady over this time period, at least when {\it averaged over the entire galaxy}.  Otherwise, if the star formation rate had (for example) been steadily decreasing during that time, we would find more RSGs relative to the number of WRs than we would expect.  Similarly, if the star formation rate had been steadily increasing, there would be fewer RSGs relative to the number of WRs than we would expect.    A short (1~Myr long) burst of star formation some 15 Myr years ago would result in an enhanced number of RSGs relative to WRs.  

For the two spirals, M31 and M33, the assumption of a constant SFR seems relatively well justified.  Using the beautiful, high spatial resolution {\it HST}  Panchromatic Hubble Andromeda Treasury (PHAT) data, \citet{2015ApJ...805..183L} finds that the SFR has been globally constant in M31 over the past 500~Myr, except for a modest (30\%) burst 50~Myr ago.   Over the past 30~Myr there may have been a gradual decline (20\%?) in the SFR (see the right-most panel in their Figure 11), but there are few accurate stellar age indicators during this time period.  Similar studies will doubtless be carried out as part of the Triangulum Extended Region (PHATTER) survey \citep{2021arXiv210101293W} for M33, but for now we are constrained by what we know from ground-based studies.  \citet{2011ApJ...738..144D} investigates the SFH of M33 over the past 250~Myr, finding that within the inner 8~kpc ($\rho<1.1$) the SFR has been constant.   

For the Magellanic Cloud the situation is more complex.   The SMC and LMC suffer interactions with each other and with the Milky Way.  As nicely summarized by \citet{2016ApJ...825...20B}, we see evidence of such past interactions in terms of the gaseous bridge between the Clouds, plus the Magellanic stream.  Such interactions are
thought to be primarily responsible for the somewhat chaotic star formation histories of the Clouds. 

 \citet{2009AJ....138.1243H} used resolved stellar photometry to determine the SFH of the LMC. Their results
show that the SFR over the past 30~Myr has increased from an average of 0.3 to 0.4~$M_\odot$ yr$^{-1}$.
This increase can be largely attributed to the birth of the 30~Dor complex about 12~Myr ago (see their Figure~17) and the Constellation~III region, born 30-50~Myr ago (their Figure~16; see also \citealt{2008PASA...25..116H}).  To further emphasize how unusual the 30~Dor region is, we note that its H$\alpha$ luminosity is $1.8\times10^{40}$ ergs s$^{-1}$ (\citealt{1999IAUS..193..418H}, and references therein), compared to the LMC's overall H$\alpha$ luminosity of $3.1\times10^{40}$ ergs s$^{-1}$ \citep{2008ApJS..178..247K}), i.e., over half of the LMC current star formation is happening in this one H\,{\sc ii} complex.  Both 30~Dor and Constellation~III are complex age-wise.  The R136 cluster at the center of 30~Dor  is a hotbed of current star formation today, containing the prototypes of the O2-3/WN5-6 objects we describe above; the suburbs of the R136 cluster (but still within the greater 30~Dor region) contain many other massive stars including classical WRs (see, e.g., \citealt{2013A&A...558A.134D}), indicating an older population than the 1-2~Myr old R136 cluster \citep{MH98}.  By contrast, Constellation~III contains (by our standards) a much older population of stars, 6-7 to 12-16~Myr \citep{1998AJ....116.1275D}.    

The SFH of the SMC is more poorly constrained.  First studied comprehensively by \citet{2004AJ....127.1531H}, the SFRs as a function of time are best shown in Figure~19 of \citet{2009AJ....138.1243H}.  At face value, the SFR has been roughly constant at about 0.15-0.30$M_\odot$ yr$^{-1}$ over the past 30~Myr, but the potential errors on this are large.  Unfortunately, the near-IR study of the SFH by \citet{2015MNRAS.449..639R} is mostly useful for lower masses (older ages), as these wavelengths are not very sensitive to recent star formation activity.  \citet{2017MNRAS.466.4540H} used UV data from SWIFT to re-examine this issue.  Their results were 
ambiguous,  in the sense that their two methods of analysis did not agree. One method indicated a 0.8~dex increase in the SFR over the past 30~Myr, and a large discrepancy with the \citet{2004AJ....127.1531H} results as shown in their Figure~17.  The other method showed a near-constant SFR over the same interval, and excellent agreement with \citet{2004AJ....127.1531H} over the past 100~Myr, as shown in their Figure~18. 

In conclusion, the SFH studies are generally supportive of our assumption that we can assume a ``steady-state" condition over the past 30~Myr, with two caveats.  First, we might do well to consider the statistics of the RSG and WR populations in the LMC by excluding both the 30~Dor and Constellation~III regions.   Secondly, we should keep in mind that the for the SMC are particularly uncertain\footnote{We note that the phenomenal success of the PHAT data in understanding the young-age SFHs of M31 \citep{2015ApJ...805..183L}, and anticipate that PHATTER will have a similar success on M33.  This is largely thanks to their inclusion of UV data, which traces the massive star populations, along with longer wavelengths that help separate intrinsic colors from the effects of reddenings.  Although the Magellanic Clouds have been partially surveyed in the UV by the Ultraviolet Imaging Telescope \citep{1998AJ....116..180P} and GALEX \citep{2014AdSpR..53..939S}, the relatively poor spatial resolutions (3\arcsec\ and 5-6\arcsec, respectively) sadly limits their suitability for such studies.  Swift has slightly better resolution (2.5\arcsec pixels), but we were unable to locate any comprehensive study of the SFHs of the LMC utilizing these data.}. We will investigate how robust our results are to this assumption in Section \ref{Sec-Sum}.

\section{The RSG/WRs Ratios as a Function of Metallicity}
\label{Sec-Ratios}

In this section we present the counts of both the RSGs and WRs in these galaxies.  We then compute the ratios and their uncertainties employing a Bayesian framework, following \citet{Fred20}.  Finally, we discuss the likely initial metallicities of these stars in preparation for comparing to the evolutionary models in Section~\ref{Sec-Models}.

\subsection{The Number of RSGs}
We list the number of RSGs in each of these galaxies in Table~\ref{tab:Numbers}.  
In counting RSGs, we choose to make cuts at three different luminosities: $\log L/L_\odot$=4.2, 4.5, and 4.8.
These values roughly correspond to initial masses of 9$M_\odot$, 12$M_\odot$, and 15$M_\odot$.  According to the single-star, rotating $Z=0.014$ models of \citet{Sylvia}, stars will
enter the RSG stage at ages of 35~Myr, 19~Myr, and 14~Myr, respectively, and remain as RSGs for about 10\% of their previous lifetimes, i.e., 3 to 1 Myr.  These ages are nearly identical for lower metallicity models (cf., the $Z=0.006$ models of P. Eggenberger et al. 2021, in prep.), and 10-20\% smaller for the non-rotating models.
These luminosity-to-initial-mass values are very approximate, as the evolutionary tracks turn nearly vertical 
during the RSG stage as the mean particle mass increases as He is converted into C and O.  Thus,  for instance, a star with an initial mass of 12$M_\odot$ enters the RSG at a $\log L/L_\odot=4.3$, but ends at a $\log L/L_\odot=4.9$ according to \citet{Sylvia}.  (See their Figure~2; for an expanded view, see Figure~8 in \citealt{M3133RSGs}.)  Weighting by time over the RSG lifetime (2.2~Myr), the average luminosity will be $\log L/L_\odot=4.54$.  (For 9$M_\odot$ and 15$M_\odot$ the time-weighted luminosities are 4.24~dex and 4.85~dex, respectively; their lifetimes as RSGs will be 2.6~Myr and 1.1~Myr, respectively.)   

The source list for the SMC comes from the Appendix of this paper.  For the LMC, the source list comes from \citet{LMCBins}.  For M31 and M33 the numbers come from \citet{M3133RSGs}, with the restrictions of $\rho\leq0.75$ and $\rho\leq 1.00$, respectively.  
(As discussed below, we also subdivide M33 into an inner, middle, and outer zone based on $\rho$, following \citealt{NeugentM33}.) 

\clearpage
\subsection{The Number of WRs}
\label{Sec-WRnumbers}

In Table~\ref{tab:Numbers} we list the number of WRs within our four galaxies based on the following surveys.

The source list of WRs for the SMC comes from \citet{MasseyWRSMC}. There are 12 WRs known, and the  recent deep survey (summarized in \citealt{FinalCensus}) failed to reveal any more.  We show their distribution in comparison with the RSGs in Figure~\ref{fig:SMCRSGWR}.

For the LMC, the source list comes from \citet{FinalCensus}, who list 154 stars.  Of these, two (BAT99-93 and BAT99-105) were ``downgraded" from O2-3If/WN6 stars to O2-3If stars, and we do not include them here. Another 10 are classified as O2-3If/WN5-6, and are mostly found in the R136 cluster at the heart of 30~Dor.  As argued above,  they are unlikely to be recognized as a WRs in the evolutionary models, and so we exclude them here as well.   However, we have retained the other two types of ``slash" stars, the Ofpe/WN9s, and the WN3/O3s, as evolutionary models should recognize both types as WRs, as discussed above.  We are left with 142 WRs.  Removing stars within 30~Dor and Constellation~III leaves 111 WRs.  In Figure~\ref{fig:LMCRSGWR} we show the distribution of the 142 WRs in comparison to the 720 RSGs with $\log L/L_\odot\geq4.5$~dex, and indicate the exclusion regions around 30~Dor and Constellation~III.  Note the large number of RSGs in the Constellation~III region with few WRs.  These RSGs are likely the result of the 12-16~Myr burst found by \citet{1998AJ....116.1275D} mentioned earlier.

For M31 our source list comes from \citet{NeugentM31}, who list 154 WRs.  One additional WR star was subsequently found by \citet{2016MNRAS.455.3453S}. (Our inspection of the survey images used by \citealt{NeugentM31} reveals that this star went undetected as it unfortunately fell in one of the narrow gaps between the CCDs in the Mosaic camera that they used.) Of these 155 WRs, 7 have $\rho>0.75$ and are not included in our counts here.  None of these are classified as O2-3I/WN5-6.  We are therefore left with 148 WRs in our comparison sample.  The distribution of RSGs and WRs are shown in Figure~\ref{fig:M31RSGWR}.  Restricting ourselves to only the regions with low extinction (i.e., with good matches between the RSGs and the LGGS), we find 69 WRs.

For M33, the counting is also a bit complicated.  \citet{NeugentM33} lists 206 WRs and six additional WR candidates.   \citet{NeugentBinaries} confirmed five of these as WRs; the sixth proved to be an Of-type star.  Subsequently, \citet{BigTable} confirmed another WR, J013344.05+304127.5 at $\rho=0.13$.  They also note that the star J013418.37+303837.0, classified as Ofpe/WN9 by \citet{NeugentM33}, is actually Var~2, one of the original Hubble-Sandage \citep{HS} stars, now called LBVs. (This confusion came about due to the previous lack of good coordinates for the Hubble-Sandage variables.)  They therefore discounted it as a WR, but here we retain its WR status as argued above.  The first two authors of the present paper are involved in a deep search for additional WRs in M33, particularly WN3/O3s; so far, they have been able to confirm one new star, J013254.12+302313.7 ($\rho=0.70$), as a WN3/O3.  Thus, the total number of known WR stars in M33 is 213; all have $\rho \leq 1.0$. 

From this M33 total we must remove several likely hydrogen-burning WRs. 
\citet{CM81} noted the similarity of several WRs in M33's brightest H\,{\sc ii} regions (NGC~604, NGC~595, and NGC~592) to the overly-bright WRs in R136 and NGC~3603 that we now classify as O2-3/WN5-6 stars.  Currently available spectra do not have the necessary signal-to-noise to accurately classify these M33 stars as O2-3/WN5-6, as the classification requires us to see absorption as well as emission.  However, it is likely that several of them fall into this category. Inspection of the \citet{NeugentM33} list allowed us to identify four stars (J013332.97+304136.1, J013311.85+303852.7, J013332.82+304146.0, and J013432.11+304705.8) whose brightness and spectral types suggest they may fall into this category.  All four are located in M33's brightest three H\,{\sc ii} regions as originally found
by \citet{CM81}\footnote{Should we also exclude these regions? NGC~604 has the second highest H$\alpha$ luminosity of any region in the Local Group, but its luminosity is $4.3\times10^{39}$ ergs s$^{-1}$ \citep{2009ApJ...699.1125R}, only about one-quarter that of 30~Dor. The integrated H$\alpha$ luminosity of M33 is $3.2\times10^{40}$ erg s$^{-1}$ \citep{2008ApJS..178..247K}, so NGC~604 accounts for only about 13\% of M33's overall H$\alpha$ luminosity, and hence its current star formation. NGC~595's H$\alpha$ luminosity is down another factor of 2.6 from NGC~604's luminosity, and NGC~592's luminosity is down a factor of 12 from that of NGC~604 \citep{2009ApJ...699.1125R}.  We suspect such regions come and go, and are more normal than such events as 30~Dor and Constellation~III, and we retain them in our sample, as we had to draw the line somewhere.}.   Thus the total number in our M33 comparison sample is 209.  We show the distribution of RSGs and WRs in Figure~\ref{fig:M33RSGWR}, and demark the inner, middle, and outer zones.

\subsection{The Ratios}

In Table~\ref{tab:Ratios} we list the RSG/WR ratios along with their associated errors. In many previous studies (see, e.g., \citealt{MJ98,MasseyOlsen,NeugentM31}) we 
have used the ratio of the observed numbers directly, and assumed that the errors were dominated by stochastic processes, i.e., that if we observed $N$ objects today, at some other time we might observe $N\pm{\sqrt{N}}$ such objects.  Thus the uncertainty $\sigma_R$ in the ratio $R=N_A/N_B$  is then computed using $\sigma_R=R\sqrt {\frac{{\sigma^2_{N_A}}    }  {N_A^2}+ \frac{{\sigma^2_{N_B}}    }  {N_B^2}       }$.  With $\sigma_A=\sqrt{N_A}$ and $\sigma_B=\sqrt{N_B}$, this simplifies to $\sigma_R=R\sqrt{N^{-1}_A+N^{-1}_B}$. However, \citet{Fred20} have recently argued one can do better than this by borrowing from the X-ray astronomy community, and
using a Bayesian framework.  This avoids such issues as when $N_B$=0 and the ratio is infinite, but finite values are also well
within a $3\sigma$ error.  Details are given in Section 3 of \citet{Fred20}, and we adopt their methodology here, and refer to the Bayesian-derived ``true" ratio $\hat{R}$.  They adopt the Markov Chain Monte Carlo package {\sc emcee} \citep{2013PASP..125..306F} to sample the posterior probability distribution
\begin{equation} 
p(\hat{R}|N_A,N_B) \propto  \int_{\hat{N}_B} d\hat{N}_B \hat{R}^{\phi - 1} \hat{N}_B^{2\phi - 1} \frac{\hat{R}^{N_A}\hat{N}_B^{(N_A+N_B)}e^{-\hat{N}_B(\hat{R}+1)}}{N_A!N_B!}
\end{equation}
where $\hat{N}_B$ is a nuisance parameter that they marginalize over corresponding to the ``true" number of stars in the denominator, and $\phi$ is a parameter to ensure good coverage, which they set to $1/2$. The $\hat{R}$ values, and their associated uncertainties, are given in Table~\ref{tab:Ratios} and will be what we use for comparing with the values derived from the evolutionary tracks.

We note that the distinction between the Bayesian-derived ``true" value (based on the expected rational numbers) and the actual observed values (based on the integer values) differ only very slightly.  The same is true for the uncertainties.  The largest difference is of course for the situation with the smallest values.  For instance, the Bayesian value for the RSG/WR ratio in the SMC for RSGs with $\log L/L_\odot\ge4.8$ is $\hat{R}=8.0^{+2.9}_{-2.0}$.  This is essentially indistinguishable from the ``classical" ratio and uncertainties calculated from the values in Table~\ref{tab:Numbers}, $8.1\pm2.5$. 

\subsection{Metallicities}

In order to compare the RSG/WR ratio to that predicted by the models as a function of  metallicity, we must know the chemical abundances of the gas out of which these massive stars were 
formed.   However, these values are not as locked in stone as often taught in introductory astronomy courses. The most common measure is that of the oxygen abundance derived from H\,{\sc ii} regions, with an assumed scaling to account for other metals. However, even the oxygen values have issues. (For an excellent review, see \citealt{2017PASP..129h2001P}.) Most commonly the nebular oxygen abundances are derived from collisionally-excited forbidden lines (such as [\ion{O}{3}]~$\lambda \lambda 4959, 5007$ and [\ion{O}{2}]~$\lambda$3727) as these lines are quite strong and easily measured in extragalactic objects.  The [\ion{S}{2}]~$\lambda6726/\lambda6831$ ratio yields the electron density
(invariably in the low-density limit), but to do the analysis, one also needs an estimate
of the electron temperature. This can be done directly if the elusive auroral lines can be measured (the so-called ``direct method"), but this is often
impossible in metal-rich regions, where the lines are very weak. For Local Group galaxies (which have negligible redshifts) there is the
additional complication that the strongest auroral line, [\ion{O}{3}] $\lambda$4363, is blended with the \ion{Hg}{1} $\lambda$4358 line, which comes
from light pollution. If the auroral lines cannot be measured, the common solution is to use $R_{23}$, one of the strong-line methods, as first defined by \citet{1979MNRAS.189...95P}.
 This derives the oxygen abundance from the ratio of the sum of [O\,{\sc ii}] and [O\,{\sc iii}] over H$\beta$, and is
calibrated with oxygen abundance via photoionization models. Both the direct and strong-lined methods depend upon an accurate correction for reddening based upon the Balmer decrement.  An additional complication is that the measurements from these collisionally-excited lines (CELs) disagree with the values derived from recombination lines (RLs) by about a factor of two, with the latter producing higher values.  This is known as the ``abundance discrepancy problem"; see, e.g., \citet{2007ApJ...670..457G}.  As we will discuss below, in some cases there is good agreement between the CEL abundances with those measured from early-type stars; in other cases, there are discrepancies.  The oxygen abundances are typically characterized as $\log$(O/H)+12, where O/H is the number ratio of oxygen atoms to hydrogen atoms.

For the Magellanic Clouds, the $\log$(O/H)+12 values appear to be well established.
The classic paper by \citet{Russell1990} list values of $8.13\pm0.04$~dex and $8.37\pm0.11$~dex for the SMC and LMC, respectively, from their study of CELs in H\,{\sc ii} regions. \citet{1998RMxAC...7..202K} used updated atomic data in their reanalysis of the older studies described by \citet{1984IAUS..108..353D}, finding values of $8.02\pm0.04$~dex and $8.37\pm0.09$~dex, respectively.  The more recent study by \citet{2017MNRAS.467.3759T} finds no evidence for radial gradients, and computes values in reasonable accord with these, $8.00\pm0.01$~dex and $8.38\pm0.01$~dex. They confirm that analysis from the RLs produce considerably higher values than those from CELs, but that the CELs are in better accord with those derived from stellar abundances of B-type stars by \citet{2009A&A...496..841H}, $7.99\pm0.21$ and $8.34\pm0.13$.  We adopt
$\log$(O/H)+12 = 8.00~dex and 8.38~dex for the SMC and LMC, respectively.  The \citet{2009ARA&A..47..481A} oxygen abundance for the sun is 8.69~dex, so this suggests metallicities that are 20\% and 50\% solar for the SMC and LMC.  The total metallicity solar metallicity $Z$ of the sun is 0.0143 \citep{2009ARA&A..47..481A}, so if we again assume that all elements scale as solar,
these correspond to $Z=0.003$ (SMC) and $Z=0.007$ (LMC). 

For the two spirals, the situation is somewhat more muddled, with conflicting estimates between the strong-lined method, the direct method, and stellar abundances.  For M31,
the classic study is that of \citet{Zaritsky1994} who used the strong-lined $R_{23}$ method; this shows a small gradient ($-0.018\pm0.006$ dex kpc$^{-1}$), and a $\log$(O/H) + 12 value of 9.0~dex at a galactocentric distance of 10~kpc, the location of the well-known star-formation ring in M31.  \citet{2012MNRAS.427.1463Z} used deep exposures to observe the auroral lines in several H\,{\sc ii} regions, allowing them to measure
temperatures (the direct method); the results show a much lower metallicity and a stronger gradient; their $\log$(O/H)+12 value at 10~kpc would be 8.5~dex or less, intermediate between that of the LMC and solar.  Another study using the strong-lined $R_{23}$ method by \citet{Sanders} found nearly identical results to that of \citet{Zaritsky1994}, with a $\log$(O/H)+12 value of 8.90~dex at 10~kpc, although they emphasize that their results depend upon the photoionization model adopted.   \citet{Venn2000} analyzed three A-type supergiants in M33; the average oxygen abundances of the two near a galactocentric distance of 10~kpc yield values of 8.7-8.8~dex
or 8.9-9.0~dex, depending upon the method used.  (We note that even though non-LTE models were used, these models failed to take into account mass-loss effects, which may be significant for A-type supergiants.)  We adopt a value of $\log$(O/H)+12=8.90~dex for M31, or about 1.7$\times$ solar.  The equivalent $Z$ is 0.030.

For M33 the situation is equally confused in terms of the nebular determinations.  The strong-line $R_{23}$ method \citep{Zaritsky1994,2010A&A...512A..63M} shows a strong gradient with galactocentric distance, with an abundance in the central region that is above solar, dropping to SMC-like in the outer region.  In contrast, the direct method (e.g., \citealt{2007A&A...470..865M,2008ApJ...675.1213R,2011ApJ...730..129B}) finds lower values in the central region and a more modest gradient.  Stellar abundances of B-type stars by \citet{2005ApJ...635..311U} agrees strongly with the higher abundances from the  $R_{23}$ method in the central region, as do the data on the distribution of WC-type subtypes \citep{NeugentM33}.  \citet{2010A&A...512A..63M} argues that the direct method likely underestimates the abundances in the central regions, as the auroral line is more easily measured in regions where the oxygen abundances is low, as oxygen is a primary source of cooling.  The abundance issue is discussed further in \citet{NeugentM33}, who adopt a two-gradient model, based upon the conclusions of \citet{2007A&A...470..865M}.  In this model, the central region ($\rho<0.25$) has a $\log$(O/H)+12 value of 8.72~dex, the middle region ($0.25\leq \rho < 0.5$) a value of 8.41~dex, and an outer region ($0.5\leq\rho\leq1.0$) of 8.29~dex.  For comparison, if we adopt the stellar results of
\citet{2005ApJ...635..311U}, we would find 8.75~dex,  8.65~dex, and 8.48~dex, respectively.  We find this agreement acceptable, but hope to improve on the situation ourselves in a future project.  The equivalent $Z$ for the metallicities we adopt (8.72, 8.41, 8.29) are 0.015, 0.008, and 0.006. We list the adopted oxygen abundances and corresponding $Z$ values in Table~\ref{tab:Ratios}.

We note that stellar evolution is primarily driven by the iron abundance as it provides the bulk of the opacity in the stellar interiors. 
In addition, it is primarily the iron abundance that drives the stellar winds and hence affects mass-loss rates among single stars. Here by using the oxygen abundance to trace the change in metallicity there could a small systematic uncertainty in our adopted metallicities. 

\section{Comparison with the Evolutionary Models}
\label{Sec-Models}

\begin{figure}
\includegraphics[width=1\textwidth]{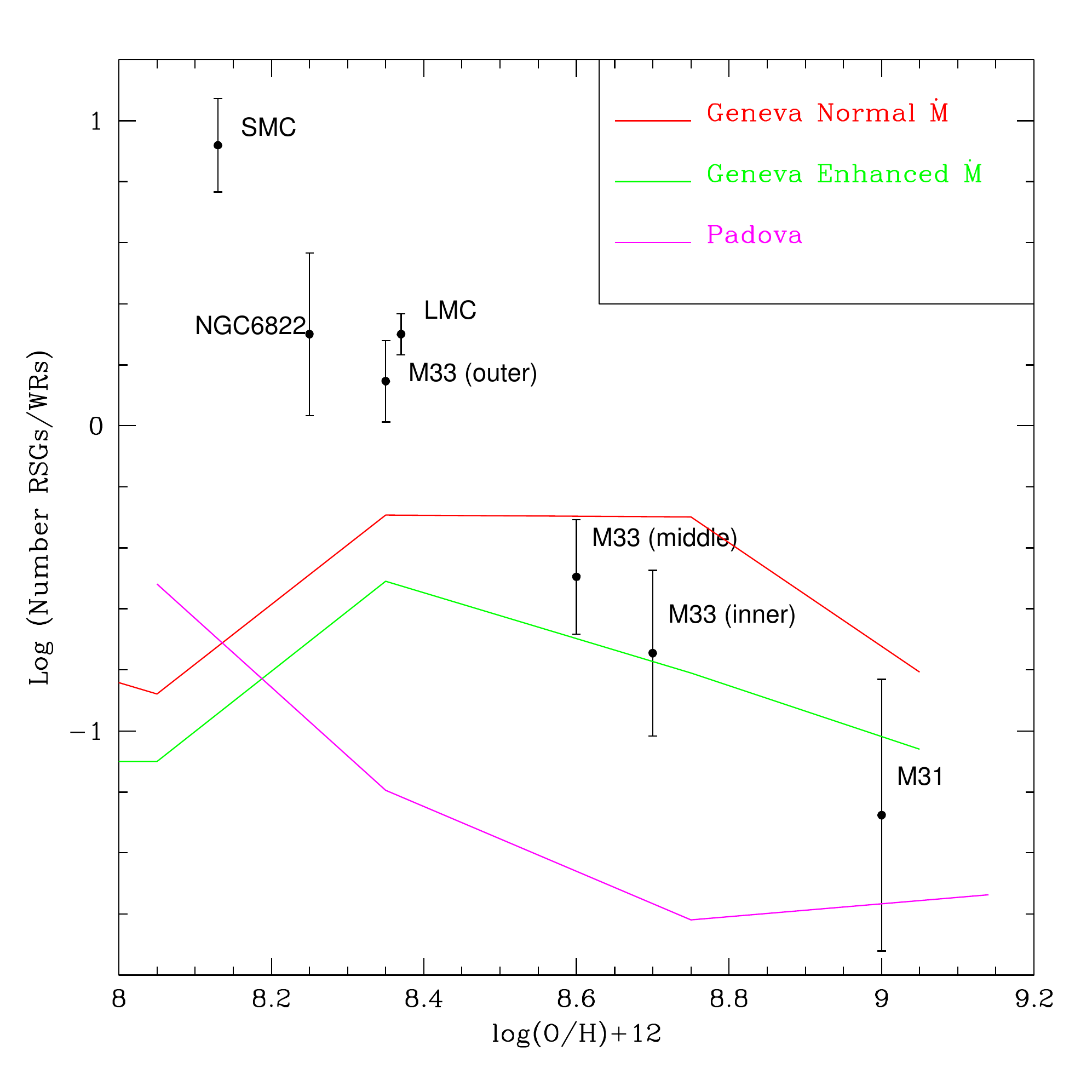}
\caption{\label{fig:araa} Original comparison of RSG/WR ratio with evolutionary models.  This figure is based on Figure~12 in \citet{MasseyARAA}, and shows the comparison between the ``observed" ratio of RSGs to WRs and that predicted by the models.  The data were the best at the time, and came from  \citet{Massey2002} and references therein.  The models used were also state-of-the-art at the time, and included mass-loss but not rotation.  The Geneva ``normal" $\dot{M}$ came from 
\citet{1992A&AS...96..269S,1993A&AS...98..523S,1993A&AS..101..415C,1993A&AS..102..339S}, while the Geneva ``enhanced" mass-loss rates came from \citet{1994A&AS..103...97M}.  The Padova evolutionary tracks came from
\citet{2000A&A...361.1023S} and references therein.}
\end{figure}

As discussed in the Introduction, the previous data showed a steeply falling RSG/WR ratio with increasing metallicity, while the (old) evolutionary models predicted a much flatter relation was expected.  A causal examination of the values in Table~\ref{tab:Ratios} shows that the new ratios calculated here no longer span the two orders of magnitudes shown in Figure~\ref{fig:araa}.  Rather, they range over less than a single order of magnitude.  Nor are they monotonic with metallicity.  How do the new data now compare with modern evolutionary models?

For this comparison, we will use the current generation of the single-star Geneva evolutionary models as well as the Binary Population and Spectral Synthesis (BPASS) evolutionary models, as described below.  We have selected these models because they have been shown to be very successful in other critical comparisons with massive star properties as described in the Introduction, and yet somehow failed to match the older RSG/WR ratio data.  However, in the hopes that present and future workers will compare their models to our data in a similar fashion, we describe in some detail here how we identify the RSG and WR phases in the evolutionary tracks in such a way that they match what we do observationally.  

The evolutionary models allow us to compute the lifetimes of RSGs and WRs as a function of initial mass.  In order to turn these into a RSG/WR number ratio, we scale these lifetimes using the assumed slope of the IMF adopting $\Gamma=-1.35$ (see Section~\ref{Sec-IMF}).   We will begin by comparing with single-star models, and then add in the complication of binarity. 

Current versions of the Geneva models are so far available at four metallicities, $Z=0.014$ \citep{Sylvia}, 0.006 (P. Eggenberger et al.\ 2021, in prep.), 0.002 \citep{2013A&A...558A.103G}, and 0.0004 \citep{2019A&A...627A..24G}.  The $Z=0.0004$ models are not useful for our purposes, as they are significantly lower in metallicity than any of the galaxies we are considering here.   At each metallicity and
initial mass there are models computed with both no initial rotation and with an initial rotation that is 40\% of the breakup speed.
A real ensemble of stars will likely have a range of rotation rates and so we have linearly interpolated between these values, i.e., a $f_{\rm rot}$ of 1 corresponds to the rotating tracks (initial rotation speed of 40\% of the breakup speed), while a $f_{\rm rot}$ of 0 corresponds to the non-rotating tracks. Note that intermediate values of $f{\rm rot}$ are approximate, and were used to simply
linearly interpolate between the results of the synthetic populations, and their values are not exact, as discussed in more
detail in Section 2.4 of \citet{Fred}.

We show the comparison between the predictions of the Geneva models with the RSG/WR ratios in Figure~\ref{fig:Geneva}.  The plots on the left show the results when we compare the results adopting the effective temperatures derived photometrically.  The agreement is far better than in previous studies (compare with Figure~\ref{fig:araa}), but the models predict too small a value at all luminosities.  However, adopting the temperature limits corrected to the spectroscopic values, as shown on the right, is very good even at the lowest RSG luminosity we consider (4.2~dex), and excellent at the higher luminosities. We can not predict what will happen at the higher metallicity corresponding to M31 ($Z=0.030$) as modern Geneva models are not yet computed for this high a metallicity, but older Geneva models showed a flattening at higher metallicities, as shown in Figure~\ref{fig:araa}.  We note that the decrease in the RSG/WR ratio with an increase of metallicity is due both to an increase in the number of WRs, and to a decrease in the number of RSGs.  For the WRs, the increase is due both to the fact that the minimum mass for forming WR stars decreases with metallicity and to the fact that the WR lifetimes increase with metallicity.  For the RSGs, the decrease is due to the fact that at higher luminosities mass loss results in the higher luminosity RSGs evolving back to the blue. 

Binary interaction is the other channel for driving the evolution of massive stars, by either donating or stripping off mass.  Although the single-star evolutionary models have enjoyed remarkable success in passing the observational tests we  (and others) have devised over the years, there is no escaping the fact that a substantial fraction of O-type stars are found in binary systems.  
The value is at least 35-40\% \citep{garmany,sana13} and may be 70\% or even higher if long-period systems are included in the analysis \citep{2015AJ....149...26A}.  The fraction of these systems that are close enough to interact during the star's lifetime is arguable, with estimates ranging up to 100\% \citep{sana12}.  A major complication is accounting for stellar mergers: as the more massive star expands during its evolution, it can engulf its neighbor, leaving behind no obvious sign (see, e.g., \citealt{SelmaMergers}). 

An essential tool for investigating the importance of binarity on massive star evolution is the BPASS evolutionary models \citep{2008MNRAS.384.1109E,2009MNRAS.400.1019E,BPASS2,2018MNRAS.479...75S}. 
These models have been used for a variety of astrophysical phenomena; recent examples include comparing the expected and observed gravitational-wave detection rates \citep{2019MNRAS.482..870E},  resolving the puzzle of an overly massive blackhole \citep{2020MNRAS.495.2786E}, and identifying a pathway for creating hydrogen-poor superluminous supernovae \citep{2021arXiv210403365S}.  Their intent, however, is primarily in exploring the effects of including binarity on stellar populations  (see, e.g., \citealt{2020MNRAS.497.2201S,Fred20}).

We show the comparison between the BPASSv2.2.1 models \citep{2018MNRAS.479...75S}\footnote{See: https://bpass.auckland.ac.nz/9.html} and the observations in Fig~\ref{fig:BPASS}.  The binary fraction $f_bin$ is again based on a linear approximation between the results BPASS single-star models and those containing a 100\% binary fraction. The plots on the left show the comparison using the photometric temperature limits,
while the plots on the right show the effect of applying the 200~K correction to bring these into accord with the spectrophotometric values.   The data suggests a modest initial binary fraction (50\%) for most metallicities.  Taken at face value, the two
extreme metallicity cases, SMC and M31, imply a much smaller initial fraction of binaries is needed.  We discuss alternative interpretations below. 

We know that binarity and rotation can have similar effects on the predictions of evolutionary models, such as the relative number of WC- and WN-type WR stars (see, e.g., \citealt{Fred20}).  Here we see a quite different behavior.  Including rotation shifts the RSG/WR ratios downwards but does not really alter the shape (Figure~\ref{fig:Geneva}).  On the other hand, adding in binarity not only lowers the curves, but also changes the shape (Figure~\ref{fig:BPASS}).  Although binarity does affect the number of RSGs, the primarily effect is in creating more WRs.  Furthermore, the WRs created by binary evolution tend to have longer lifetimes than those created by single-star evolution since they have smaller masses.  Even for higher mass stars that would have become a WR star without binary evolution, binary interactions will cause the star to become a WR star sooner, so more of its evolution is during a WR phase.  The relative importance of the binary channel becomes less important at higher metallicities, as the higher mass-loss rates that come with larger metallicities lowers the mass limit for becoming a WR.  These lower-mass WRs have longer lifetimes during the WR phase, and so the net effect of binarity is to  flatten the RSG/WR ratio as a function of metallicity.

\clearpage
\section{Summary and Discussion}
\label{Sec-Sum}

We have determined the RSG/WR ratio for four Local Group galaxies, covering an order of magnitude in metallicity, from $Z=0.003$ to $Z=0.030$.  We find that there is only a modest variation in this quantity with metallicity, contrary to the original
suggestion by \citet{Maeder80}\footnote{What, then, about \citet{Maeder80}'s assertion that the RSG/WR ratio varies by two orders of magnitudes within 2 ~kpc of the sun?  This is a subject worthy of reexamination now with {\it Gaia} distances and with our improved knowledge of the WR and RSG content locally, but is beyond the scope of the present paper.} and earlier estimates (e.g., \citealt{MasseyARAA}).   However, in the larger sense, \citet{Maeder80} has certainly proven to be right, in that the RSG/WR ratios provide a very exacting test of modern stellar evolution theory.

In making the comparison with the Geneva and BPASS evolutionary models, we have been careful to match what we are using to define the RSG and WR phases in the models with what we are using for identifying these stars observationally. For the WRs, this means ignoring the handful of WRs that are likely still hydrogen-burning classified as O2-3If/WN5-6 stars and found in the largest H\,{\sc ii} regions. For the RSGs, we have used both a luminosity limit in defining the number of RSGs, and used the same temperature criteria in the models as we did observationally.  When we do this, both the Geneva single-star models and the BPASS binary models do a good job of matching the observed ratios, as long as we adopt the temperature criteria corresponding to the spectroscopic effective temperatures rather than that derived from photometry.  This is consistent with the recent analysis by \citet{Ironman}, but runs contrary to the conclusion by \citet{Davies2013}.  When dealing with stars with extended atmospheres, such as RSGs, the definitions of radius and effective temperatures may differ between evolution interior models and atmospheric models.  The fact that the spectroscopic temperatures agree better with the placements of stars in the HRD relative to the tracks is therefore encouraging.

Another factor to consider is that what we identify as RSGs and WR stars in the evolutionary models is, to some extent, uncertain. For RSGs, the model stars become large and cool, but it is unclear if their theoretical surface temperatures mean exactly the same thing as what we measure observationally as the temperatures of real RSGs. It is possible that we are missing some of the RSGs at lower metallicity because they remain hotter than the temperature limits imposed here. The correct boundary conditions for RSGs are also unknown, and the BPASS and Geneva models both use different methods to evaluate the surface of the RSG phase of evolution.

For WR stars, the luminosities of the evolutionary models mean the same thing as what we derive observationally from quantitative analysis of the spectra, but the meaning of their theoretical surface temperatures are more difficult to relate to
what we derive from such studies.  Here the fact the stars must be hydrogen free allows them to be easily identified. Uncertainty occurs when we begin to consider the lowest luminosity for a star to be a WR star. A low mass and luminosity helium star will still eventually undergo core-collapse but may not have a high enough luminosity to drive an optically thick winds and produce the characteristic emission WR lines, as discussed by \citet{2018A&A...615A..78G}. At solar metallicity this luminosity limit is $\log(L/L_{\odot}) \sim 4.9$. Recently \citet{2020A&A...634A..79S} suggested that this limit increases at lower metallicities due to the changing opacity of the stellar interiors making it more difficult to drive stellar winds. This has not been taken into account in this study; however, \citet{2020MNRAS.497.2201S} showed that including such a limit did decrease the number of stars observed as WR stars in a stellar population which would improve agreement here between the BPASS models and the SMC numbers (see their Figure 4).

We note this is only important for binary populations because the single-star Geneva models only have the highest mass stars becoming WR stars. With a binary population a much wider range of stellar masses become WR stars and without an accurate minimum WR star luminosity imposed, these models over-predict the number of WR stars at lower metallicities. If such a varying limit is correct, then there may be a number of products of binary interactions in low metallicity environments that are difficult to identify.

How robust is our result to variations in the star formation rates? As emphasized earlier (Section~\ref{Sec-IMF}), because of the difference in ages between RSGs and WRs, the expected 
RSG/WR ratio depends upon our assumptions about the constancy of the SFR.   Let us briefly explore this, using as our example the BPASS models after adopting the spectroscopic temperatures and $\log L/L_\odot > 4.5$ for the RSGs.  
Fig~\ref{fig:20} shows the effect of a steadily increasing or decreasing SFR over the past 40~Myr.  For a 30\% change in the SFR, the results are indistinguishable from the assumption of a constant SFR.

Perhaps a more realistic approximation for changes in the SFH is shown in the upper panel of Figure~\ref{fig:yarn}.   We simulated changing the SFHs using a Gaussian process, such that star formation is correlated on $\sim$5 Myr timescales, and varies with an amplitude of 10-40\%.  We used 200 random simulated SFHs; time is logarithmic to show the smooth variations on Myr timescales, but are stochastic on longer timescales.  When we applied this to the models, we obtained the comparison shown in the bottom panel of Figure~\ref{fig:yarn}.  Although it broadens the expected ratios at a given metallicity, it does not change the results.

Although the agreement with the models is very good, there are things we can learn from the discrepancies.  First, in terms of the Geneva models, we find that the agreement is best when using an intermediate initial rotation rate.   They adopted 40\% of the breakup speed as their initial rotation based upon the observational work of \citet{2006ApJ...648..580H} and \citet{2010ApJ...722..605H}.  But we actually expect there to be a distribution of initial rotation velocities as stars are formed.  The agreement with the RSG/WR ratios suggest an intermediate value, but of course these models do not include the effect of binarity.  We will note that based on a more limited metallicity range than is currently available, \citet{NeugentM31} also found that the relative number of WC and WN stars fell between the predictions of the rotating and non-rotating models (see their Figure~10) although this should be reexamined now that we have more complete data.

In terms of the BPASS models, we see really good agreement for most of the points at an intermediate initial binary fraction (about 60-70\%), except for the low and high metallicity examples (SMC and M31), which suggest a low binary frequency. Metallicity could plausibly affect the binary fraction in stellar populations through modifying the opacity of proto-stellar disks and hence the fragmentation of molecular clouds. This is challenging for hydrodynamical simulation to evaluate, and outcomes remain mixed in their predictions for any trend in binary fraction of massive stars with metallicity (see, e.g., \citealt{2020MNRAS.491.5158T} for discussion).  Or, it could be that (for the SMC), as noted above our luminosity limit for when to include WR stars is too low.

\citet{M31M33RSGBins} found that the intrinsic binary fraction of RSGs is correlated with metallicity, with a higher binary fraction at larger metallicities; see also \citet{2018ApJ...867..125D}. This is the reverse of the behavior seen in lower mass Solar-type stars, which are more likely to lie in binaries at low metallicities \citep{2019ApJ...875...61M,2020MNRAS.499.1607M} and also show a dependence on $\alpha$-enhancement, hinting at the complexity of disk fragmentation physics. This observed trend in RSGs could explain the SMC being consistent with a low binary fraction, but not the M31. Note, too, that the observed binary fraction of WRs appear to be independent of metallicity \citep{2001MNRAS.324...18B,2003MNRAS.338..360F,2003MNRAS.338.1025F,NeugentBinaries}, although past mergers may could affect those statistics. 

Low apparent binary fractions at both ends of the metallicity range probed could hint that more than one aspect of the population is changing, with a different metallicity dependency. In this paper and in the context of the BPASS evolutionary models used, the term binary fraction is a simplification of a more complex parameter space and does not account for the variation in binary properties. In addition to an initial binary fraction, BPASS must also assume initial distributions in binary separation and mass ratio between the components. In BPASS v2.2 (used here) these are based on the empirical distributions of \citet{2017ApJS..230...15M}, which have been calibrated on local stellar populations but have not been evaluated at high or low metallicities.  A metallicity-dependent change in the parameter distributions favoring slightly wider binaries, or those with lower mass ratios, would decrease the probability of interaction between binary components (rendering the stars effectively-single) and thus have the same apparent effect on stellar number counts as a lower initial binary fraction. Further observations and modeling of binary populations in both high and metallicity environments are required to determine whether such an explanation may plausibly explain the results for M31.

It should also be recalled in passing that while the BPASS models model binary interactions in detail, they do not include the evolutionary effects of higher order multiple systems or of ongoing stellar rotation, which we know are also important, particularly at low metallicities. 

Observationally, we can also make improvements.   For instance, \citet{2016MNRAS.455.3453S} discovered WR star in M31 that was not found as part of the \citet{NeugentM31} study.  The star is described as ``heavily reddened," although no actual numbers are given.  Is this star indicative of a missing population of WRs? We noted earlier that this WR was missing from the \citet{NeugentM31} survey only because it fell within a gap between the mosaic CCD camera they used for their survey, but the question is still worth investigating.   \citet{2019Galax...7...74N} argues why a missing population of WRs is highly unlikely, but prompted by the current results, the first two authors plan a deeper survey in several spots in M31 to check out this possibility. (We note again that the RSG/WR ratio found using the entire sample, and that found for the regions with lowest extinction, agree within the uncertainties.)   Another improvement observationally would be to include additional galaxies, but here the options are rather limited.  It would be practical to extend this work to include the galaxy NGC~6822, which  has low metallicity ($\log$(O/H)+12=8.25 or $Z=0.005$, \citealt{1980MNRAS.193..219P}) and is known to contain four WRs \citep{MJ98}.  \citet{2021A&A...647A.167Y} have identified RSGs in this galaxy using a unique method, but no studies have yet determined luminosities or temperatures for these objects; such a study involving several of the present authors is underway (T. Dimitrova et al.\ 2021, in prep.) However, other Local Group galaxies either have too few WRs to be statistically useful, or else have other issues.  For instance,   IC~1613 has a very
attractive low metallicity ($\log$(O/H)+12=7.80 or $Z=0.0018$, \citealt{2007ApJ...671.2028B}), but contains just a lone WR, of WO type. \citet{2019Galax...7...74N} cites this as an example of a WR that has almost certainly been formed by binary evolution, given the low metallicity environment and the resulting expectations that stellar winds would be ineffective\footnote{\citet{2021ApJ...909..113L} identifies an IR source in IC1613, SPIRITS 14bqe, as a possible dust-forming WC binary.}.  In contrast, the Local Group galaxy IC~10 contains several dozen WRs \citet{1995AJ....109.2470M,2002ApJ...580L..35M,2017MNRAS.472.4618T}, and has a metallicity similar to that of NGC~6822 \citep{1990ApJ...363..142G}.  A preliminary accounting of IC10's RSG population is given by  \citet{2020RNAAS...4..107D}.  However, it is clearly undergoing a galaxy-wide starburst phase (e.g., \citealt{1995AJ....109.2470M, vandenbergh2000}). Thus its RSG/WR ratio would not be a very useful test of models without a clearer understanding of the recent evolution of its SFR. 

Using the RSG/WR value as a test of stellar evolution models is particularly challenging because of the difference in mass ranges leading to the RSG and WR stages.  We have previously explored the use of the relative number of WC and WN stars both for the Geneva and BPASS models (see, e.g., \citealt{MasseyARAA,NeugentM31,BPASS2,2020MNRAS.497.2201S}).  However, one of the most useful observational tests has been the hardest to realize, namely the an accurate measure of the relative number of blue and red supergiants, or, more specifically, the number of O-type stars to WRs (see, in particular, predictions in \citealt{2018ApJ...867..125D,Fred20,2020MNRAS.497.2201S}).  The difficulties of a meaningful census of have been discussed elsewhere (see, e.g., \citealt{MasseyBridge,2017RSPTA.37560267M}), but observational progress continues to be made on that front.

In summary, the long-standing problem of the mismatch of the RSG/WR ratio and the evolutionary models has been resolved satisfactorily. The difference between this study
and previous comparisons has been in the improved observational material rather than
changes in the evolutionary models. We have utilized  the complete surveys for WRs and RSGs that have become available over the years, and also made use of  comprehensive evolutionary models that include rotation (Geneva) and binarity (BPASS).   We note that other modern evolutionary models are available, such as the MESA Isochrones and Stellar Models \citep{2016ApJS..222....8D,2016ApJ...823..102C,2011ApJS..192....3P,2013ApJS..208....4P,2015ApJS..220...15P}, the FRANEC models \citep{2013ApJ...764...21C,2018ApJS..237...13L} and the long-awaited Bonn evolutionary models \citep{2020arXiv200408203S}.  All of these include rotation, but not binarity. In general, we are unaware of detailed comparisons that have been made with these models with observations of massive stars, such have been carried out over the past decade for the Geneva and BPASS models.   We would encourage their adherents to test these models against the RSG/WR ratios using the same methodology we have described here.

\begin{acknowledgements}

The lead author, PM, would like to dedicate this paper to his former thesis advisor, Peter S. Conti, who recently passed away, and who would have enjoyed
this study; it was also he who excitedly called the \citet{Maeder80} paper to PM's attention when it first appeared.

Lowell Observatory sits at the base of mountains sacred to tribes throughout the region. We honor their past, present, and future generations, who have lived here for millennia and will forever call this place home.
 Similarly, KFN, TZD-W, and EML would like to acknowledge that they work  on the traditional land of the first people of Seattle, the Duwamish People past and present, and honor with gratitude the land itself and the Duwamish Tribe. 
 
We are indebted to Georges Meynet for his thoughtful comments on an early draft of the manuscript.  An anonymous referee also provided comments that helped us improve the paper.  We also thank Maria Drout for help with the analysis of Gaia data in connection with earlier papers.
This work was partially supported by the National Science Foundation (NSF) through AST-1612874.   Although we do not include any new observations in the present paper, we wish to acknowledge that the two decades of work that have led to the present improvements were made possible thanks to the excellent observing facilities at Las Campanas Observatory (Magellanic Cloud WR surveys and spectroscopic followups), the Kitt Peak National Observatory (M31/M33 WR surveys), 
the MMT (M31/M33 WR spectroscopic followups),  the United Kingdom Infrared Telescope and the Canada France Hawaii Telescope (M31/M33 RSG surveys).   We are grateful to the generous and continuous support by the time allocation committees of the University of Arizona Observatories and the National Optical Astronomy Observatories.  Similarly our knowledge of the RSG content of the Magellanic Clouds would not be possible without the data products from the Two Micron All Sky Survey, which is a joint project of the University of Massachusetts and the Infrared Processing and Analysis Center/California Institute of Technology, funded by the National Aeronautics and Space Administration and the NSF.

Some of our illustrations were based on photographic data obtained using Oschin Schmidt Telescope on Palomar Mountain.  The Palomar Observatory Sky Survey was funded by the National Geographic Society.  The Oschin Schmidt Telescope is
operated by the California Institute of Technology and Palomar
Observatory.  The plates were processed into the present compressed
digital format with their permission.  The Digitized Sky Survey was
produced at the Space Telescope Science Institute under U. S. Government grant NAG W-2166.  

This work has made use of data from the European Space Agency (ESA) mission
{\it Gaia} (\url{https://www.cosmos.esa.int/gaia}), processed by the Gaia
Data Processing and Analysis Consortium (DPAC,
\url{https://www.cosmos.esa.int/web/gaia/dpac/consortium}). Funding for the DPAC
has been provided by national institutions, in particular the institutions
participating in the {\it Gaia} Multilateral Agreement.

\end{acknowledgements}

\facilities{Swope (SITe No.\ 3 imaging CCD, E2V imaging CCD), Magellan:Baade (MagE spectrograph), Mayall (Mosaic-1 wide-field camera), MMT (Hectospec multi-fiber spectrograph, Blue Channel), UKIRT (WFCam NIR wide-field camera), CFHT (WIRCam NIR wide-field camera), CTIO:2MASS}

\clearpage

\appendix
\section{Source List of SMC RSGs}

\begin{figure}
\includegraphics[width=1\textwidth]{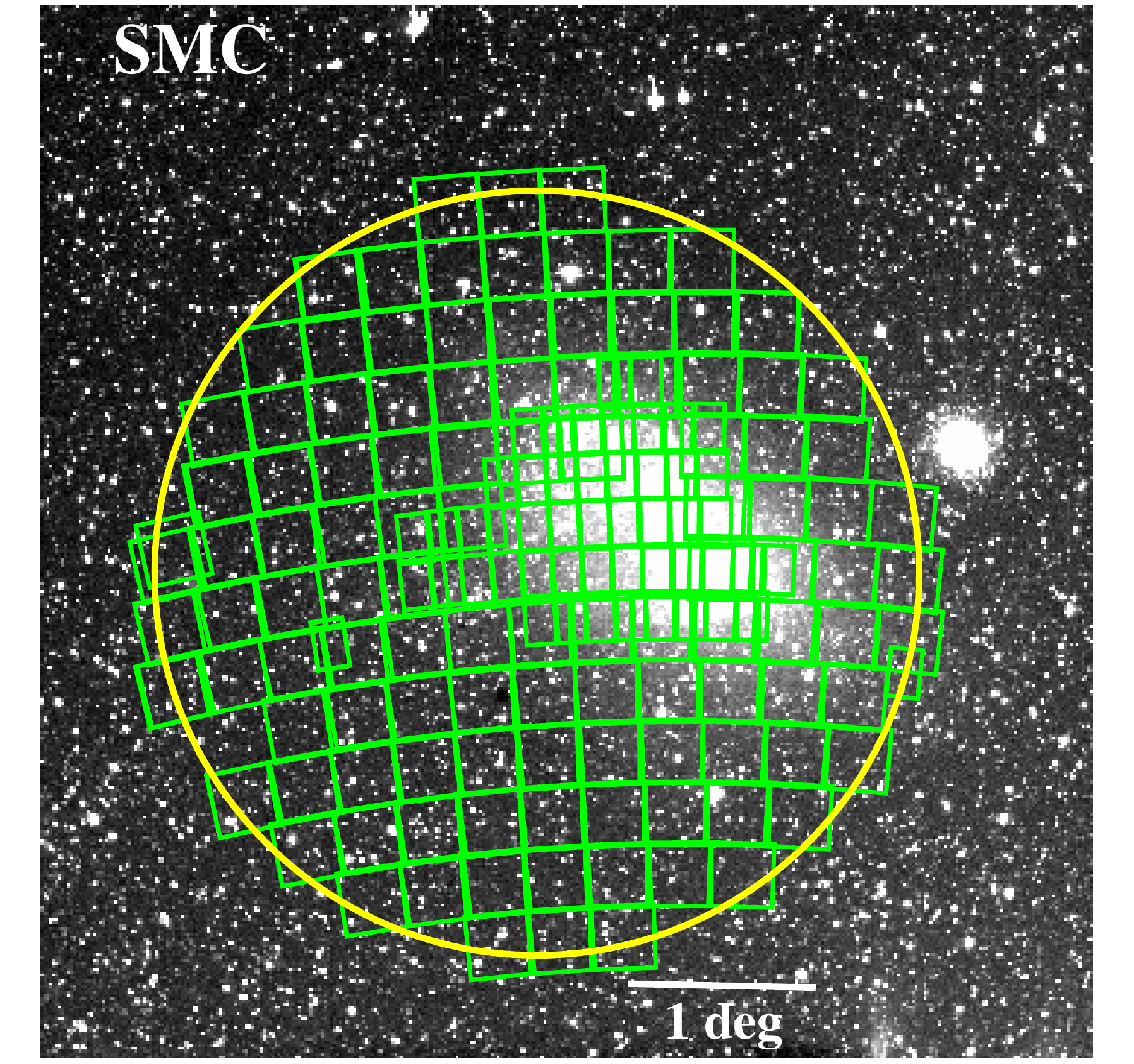}
\caption{\label{fig:smc} SMC Survey Area.  The yellow circle denotes the survey region for the RSGs (this paper), and the green boxes denote the survey region for the WRs \citep{FinalCensus}. The underlying R-band image of the SMC is from \citet{ParkingLot} and was kindly provided by G. Bothun.}
\end{figure}

\begin{figure}
\includegraphics[width=1.05\textwidth]{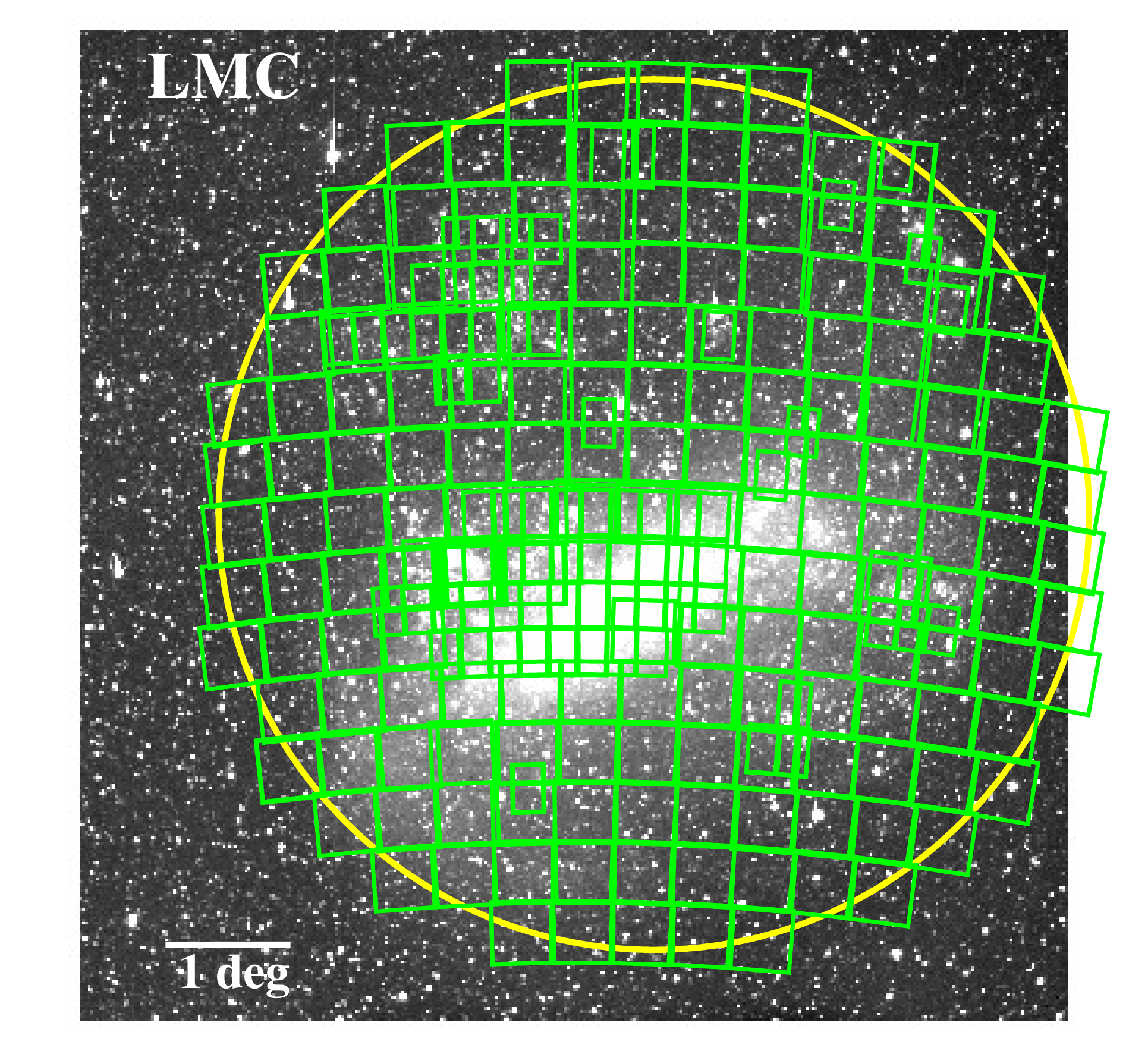}
\caption{\label{fig:lmc} LMC Survey Area.  The yellow circle denotes the survey region for the RSGs \citep{LMCBins}, and the green boxes denote the survey region for the WRs \citep{FinalCensus}. The underlying R-band image of the LMC is from \citet{ParkingLot} and was kindly provided by G. Bothun.}
\end{figure}

\begin{figure}
\includegraphics[width=1.05\textwidth]{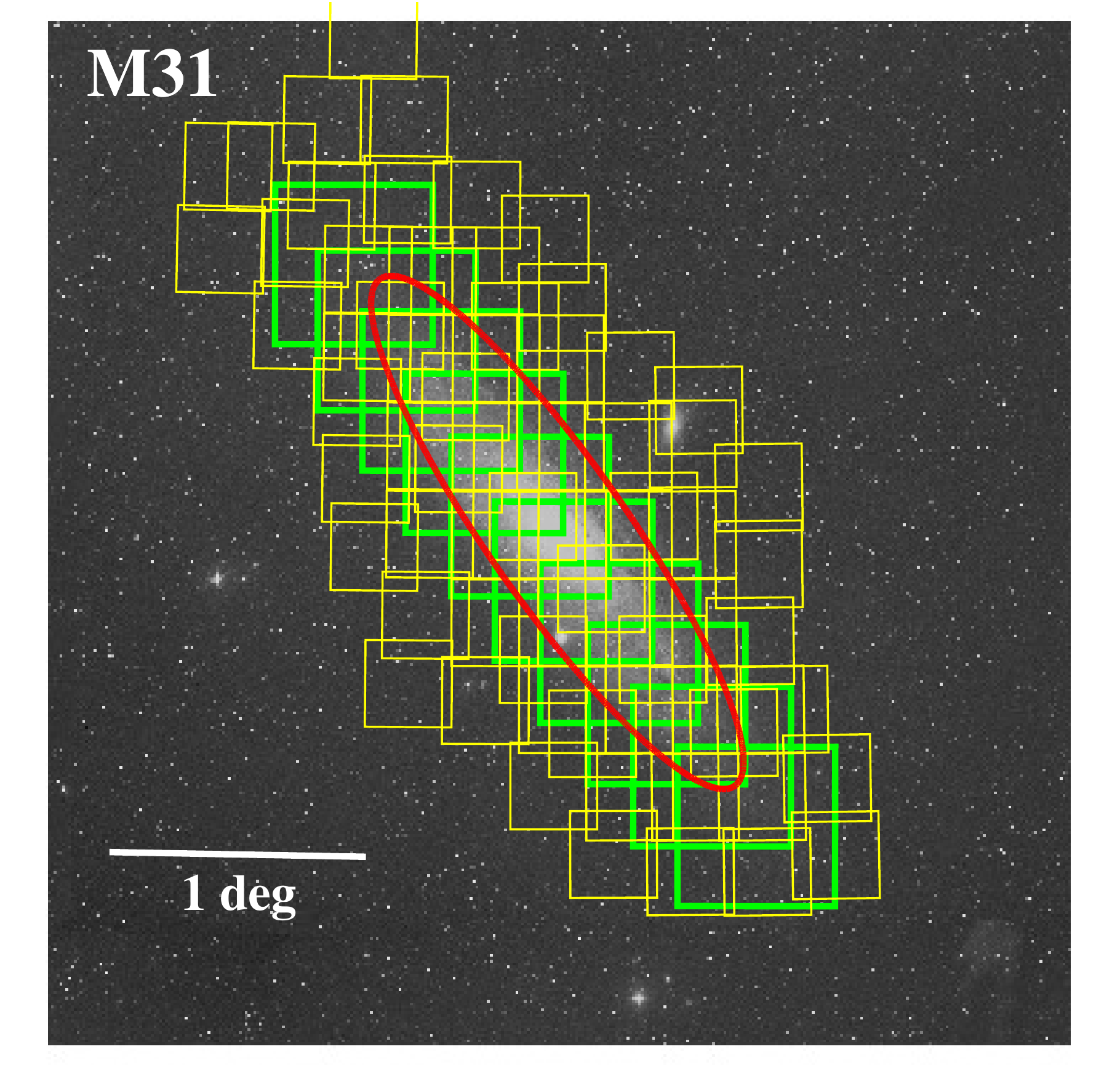}
\caption{\label{fig:m31} M31 Survey Area.  The yellow boxes denote the survey region for the RSGs \citep{M3133RSGs}, and the green boxes denote the survey region for the WRs \citep{NeugentM31}.  The red oval denotes a galactocentric radius $\rho=0.75$. The underlying V-band image covers an area 4\fdg0 $\times$4\fdg0 on a side and was provided by NASA's SkyView server from the digital sky survey. This figure is adapted from \citet{M3133RSGs}.}
\end{figure}

\begin{figure}
\includegraphics[width=1\textwidth]{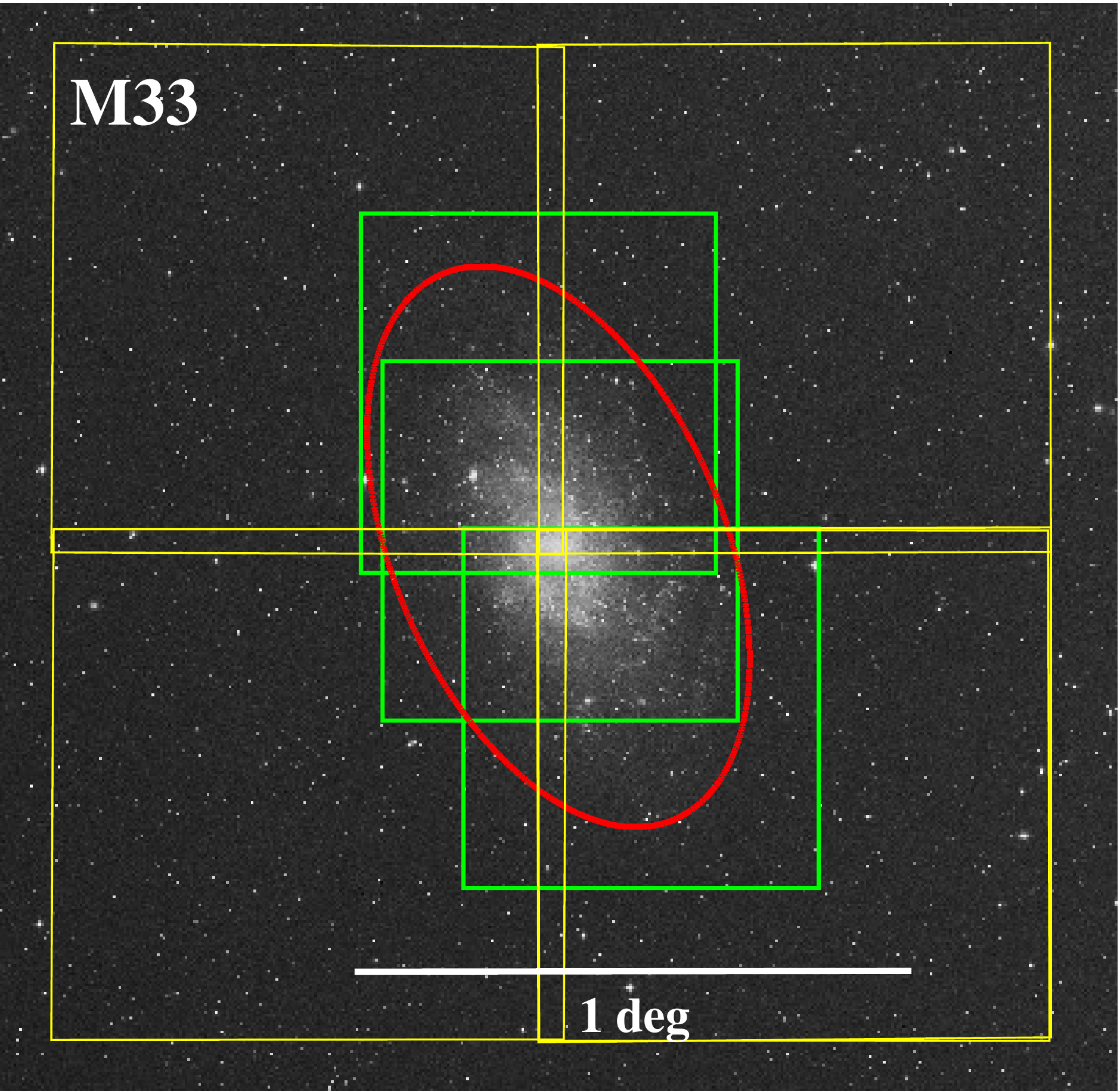}
\caption{\label{fig:m33} M33 Survey Area.  The yellow boxes denote the survey region for the RSGs \citep{M3133RSGs}, and the green boxes denote the survey region for the WRs \citep{NeugentM33}.  The red oval denotes a galactocentric radius $\rho=1.00$. The underlying V-band image covers an area 2\fdg0 $\times$2\fdg0 on a side and was provided by NASA's SkyView server from the digital sky survey. This figure is adapted from \citet{M3133RSGs}.}
\end{figure}

\begin{figure}
\includegraphics[width=1\textwidth]{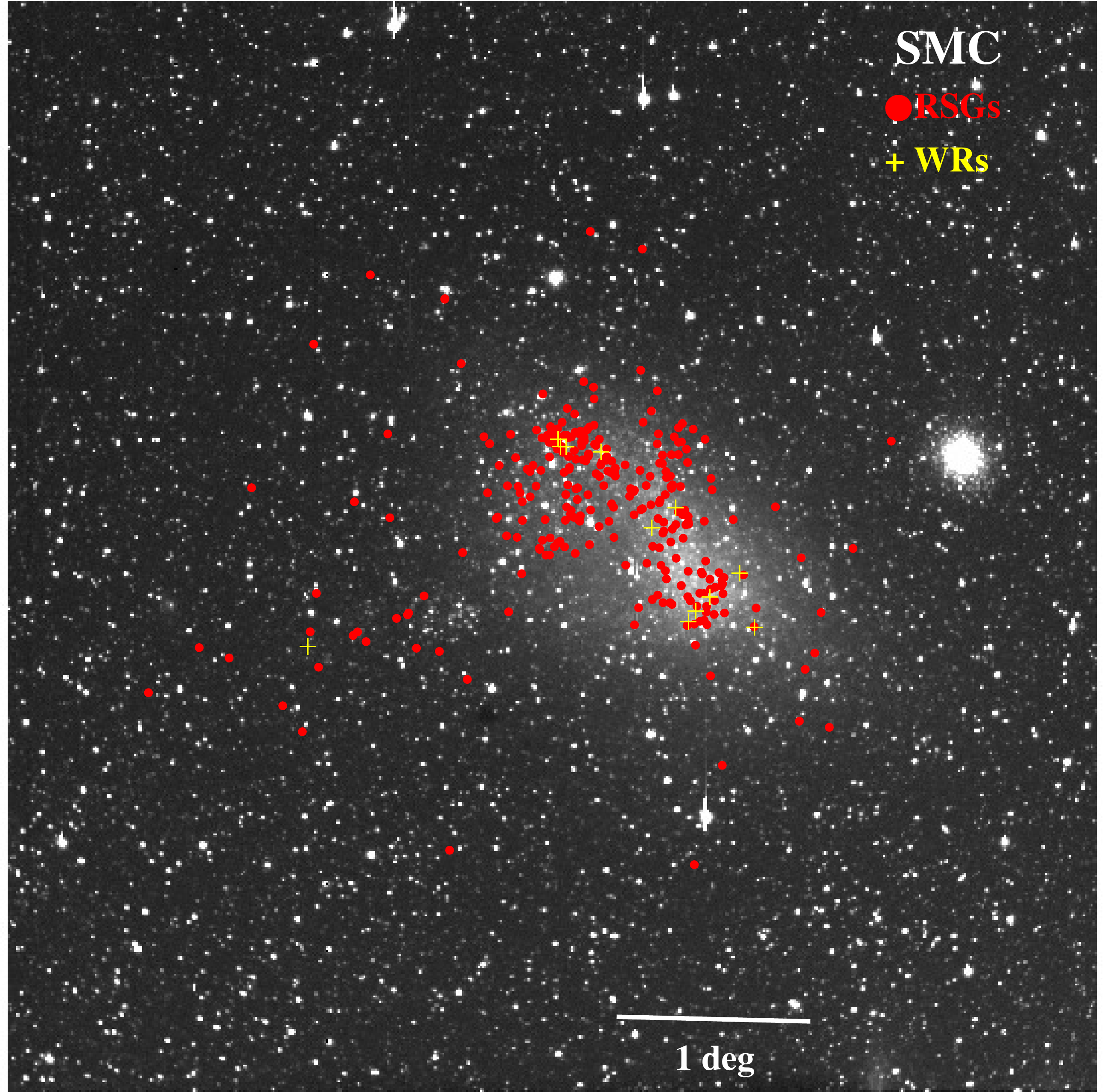}
\caption{\label{fig:SMCRSGWR} Distribution of RSGs (red points) and WRs (yellow plus signs) in the SMC.  The underlying R-band image of the SMC is from \citet{ParkingLot} and was kindly provided by G. Bothun.}
\end{figure}

\begin{figure}
\includegraphics[width=1.1\textwidth]{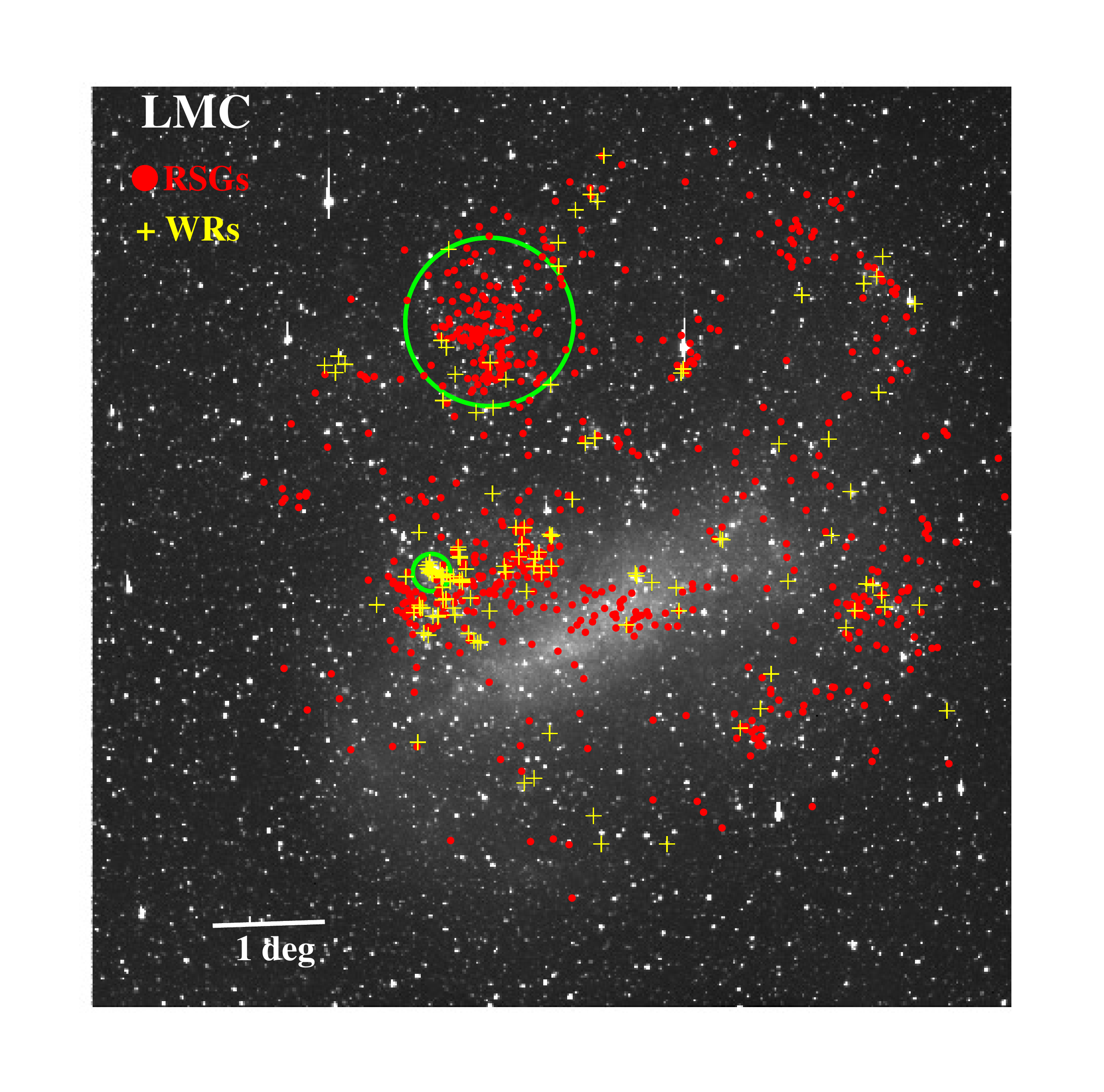}
\caption{\label{fig:LMCRSGWR} Distribution of RSGs (red points) and WRs (yellow plus signs) in the LMC.   The two green circles denote the two exclusion regions for our sample, the upper being Constellation~III, and the lower being 30~Dor. The underlying R-band image of the LMC is from \citet{ParkingLot} and was kindly provided by G. Bothun.}
\end{figure}

\begin{figure}
\includegraphics[width=1\textwidth]{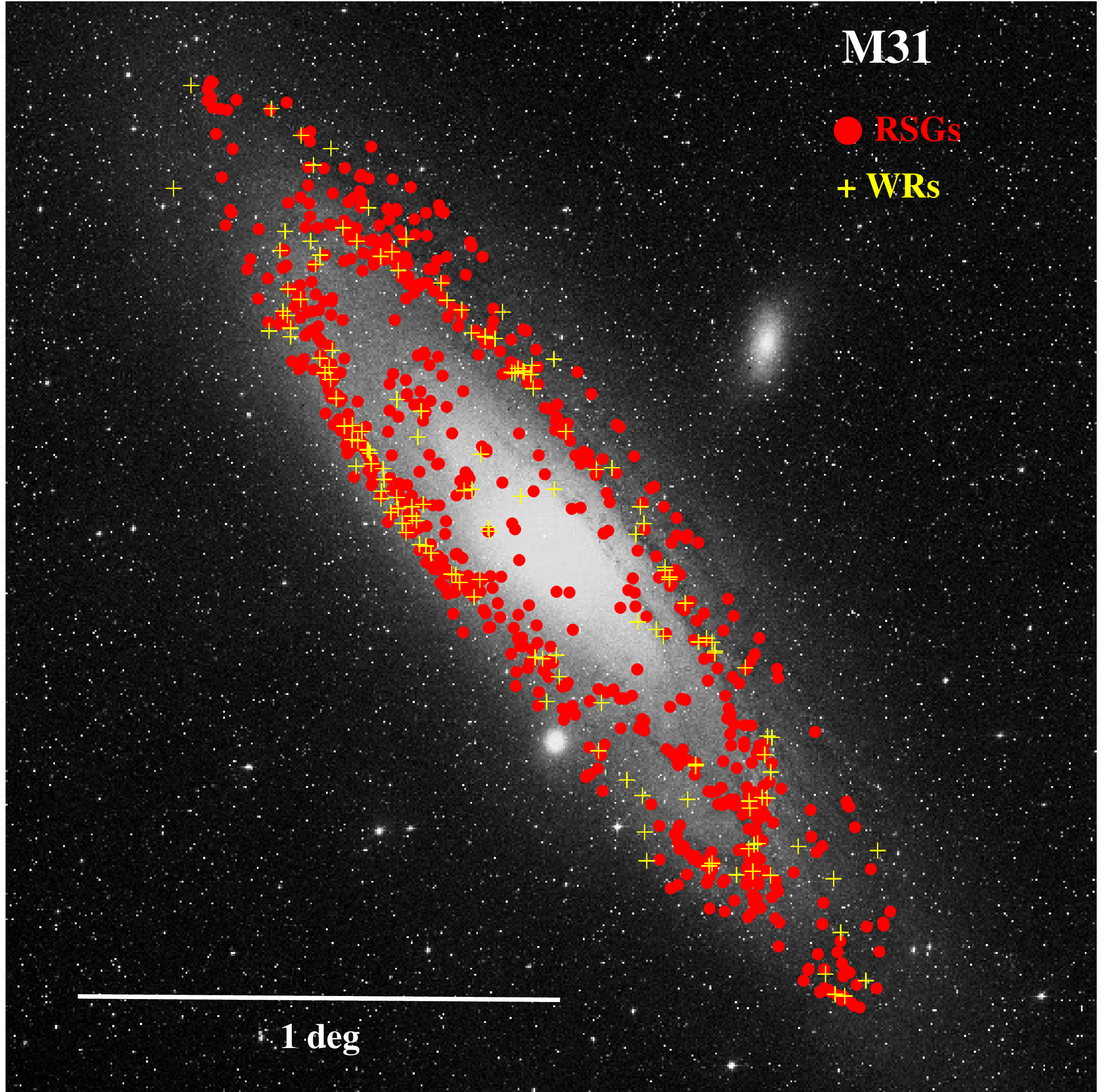}
\caption{\label{fig:M31RSGWR} Distribution of RSGs (red points) and WRs (yellow plus signs) in M31. The underlying V-band image covers an area 2\fdg25 $\times$2\fdg25 on a side and was provided by NASA's SkyView server from the digital sky survey.}
\end{figure}

\begin{figure}
\includegraphics[width=1\textwidth]{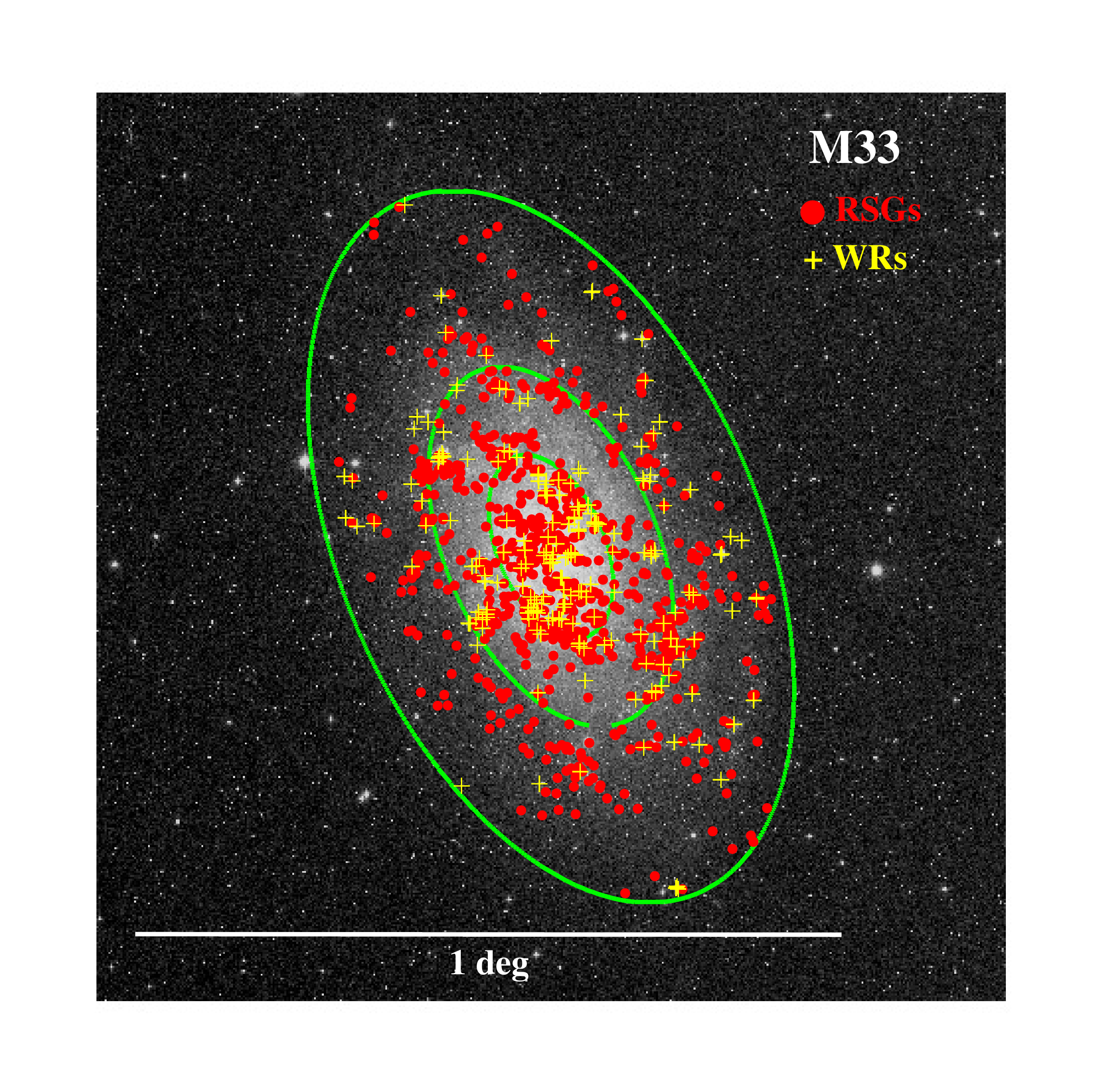}
\caption{\label{fig:M33RSGWR} Distribution of RSGs (red points) and WRs (yellow plus signs) in M33. The three green ovals denote galactocentric distances of $\rho$=0.25, 0.5, and 1.0 within the plane of the galaxy.  The underlying V-band image covers an area 1\fdg25 $\times$1\fdg25 on a side and was provided by NASA's SkyView server from the digital sky survey.}
\end{figure}

\begin{figure}
\includegraphics[width=0.55\textwidth]{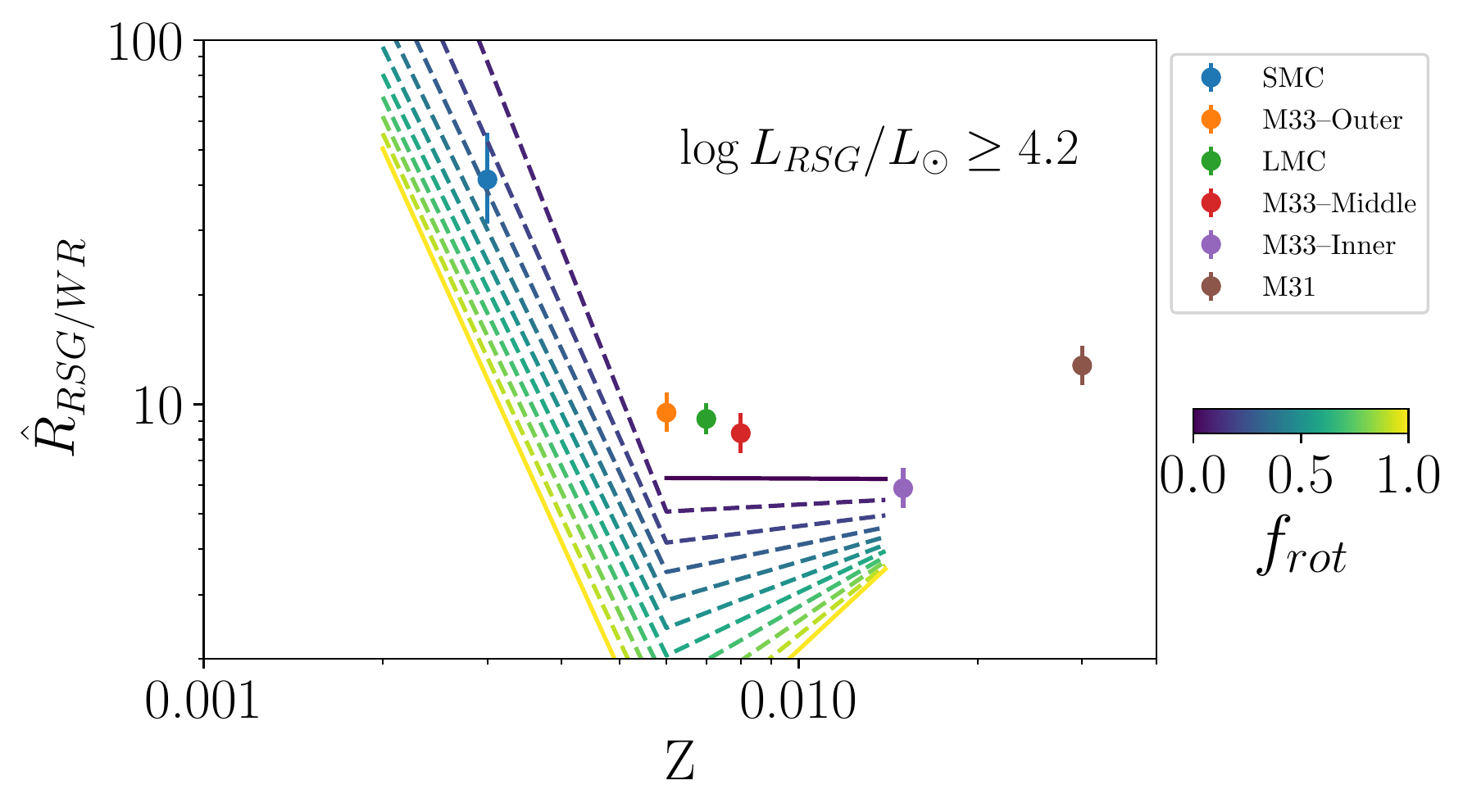}
\includegraphics[width=0.55\textwidth]{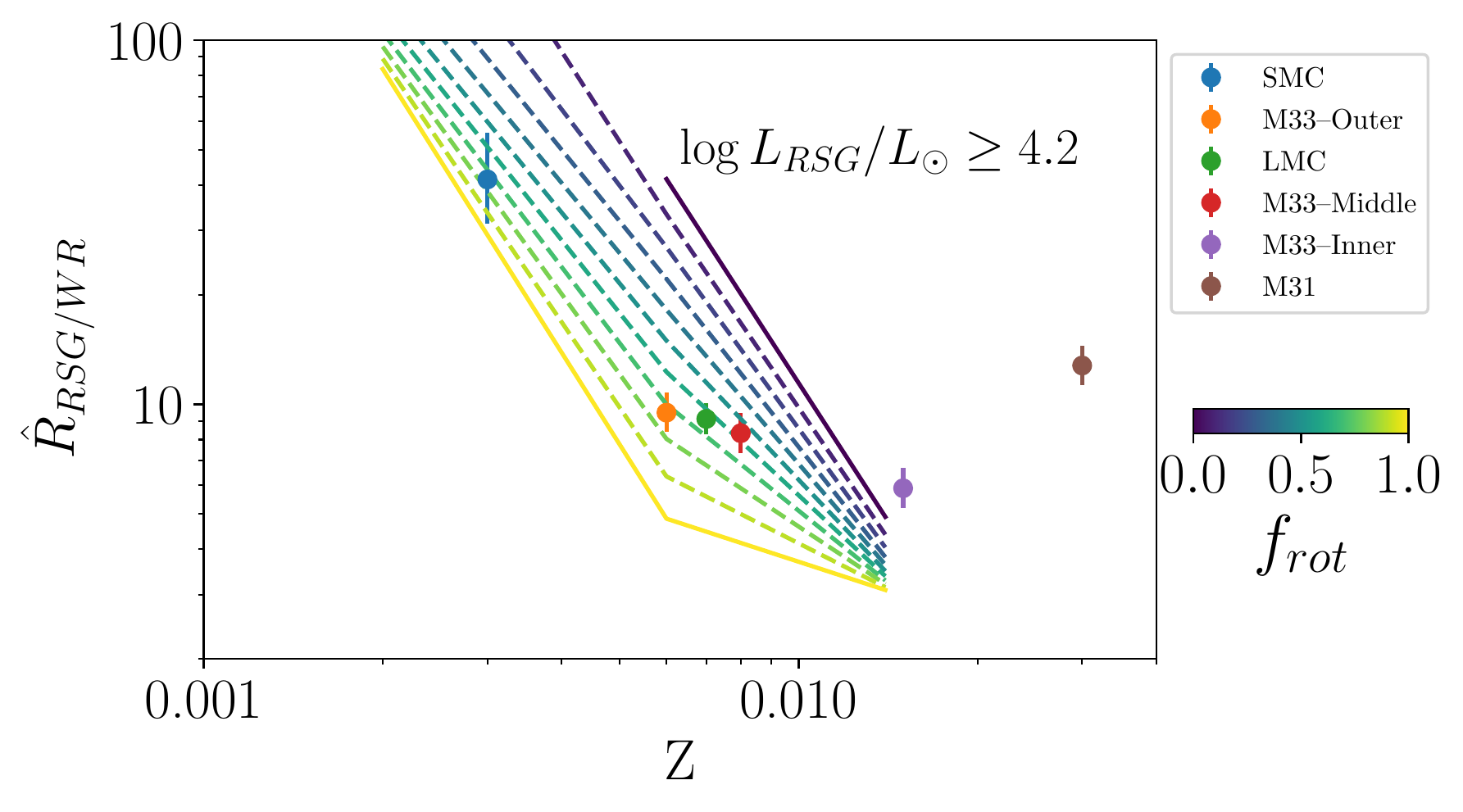}
\includegraphics[width=0.55\textwidth]{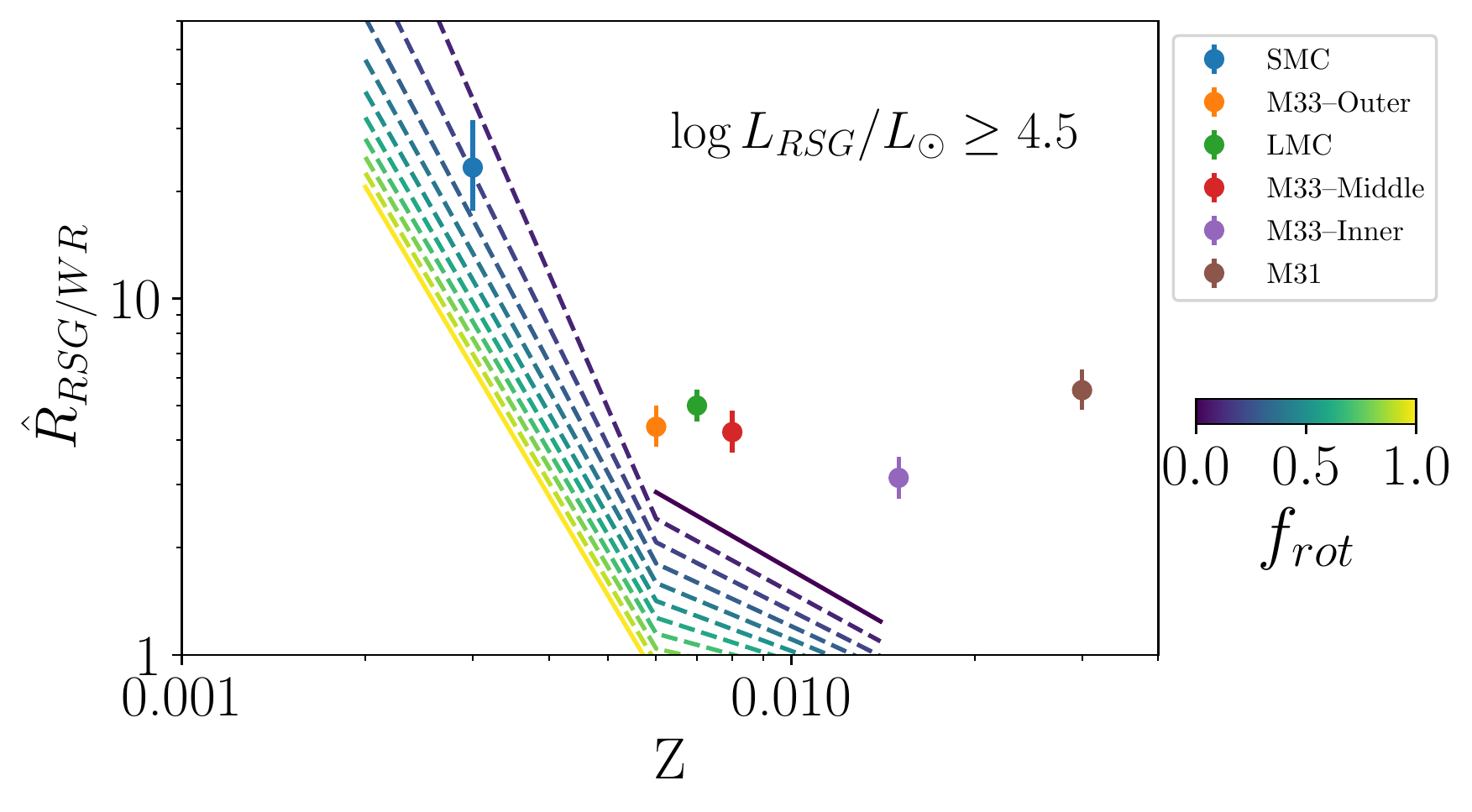}
\includegraphics[width=0.55\textwidth]{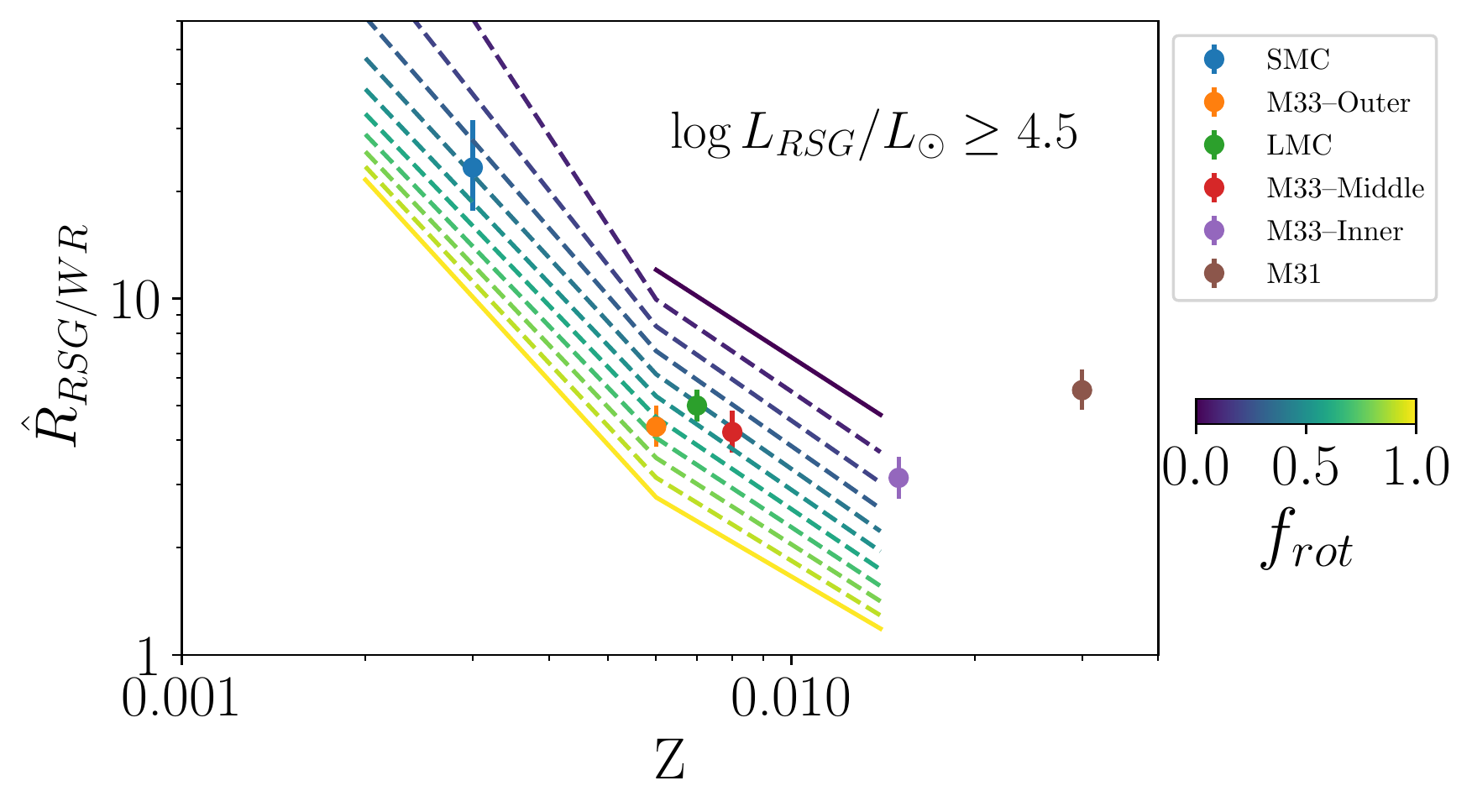}
\includegraphics[width=0.55\textwidth]{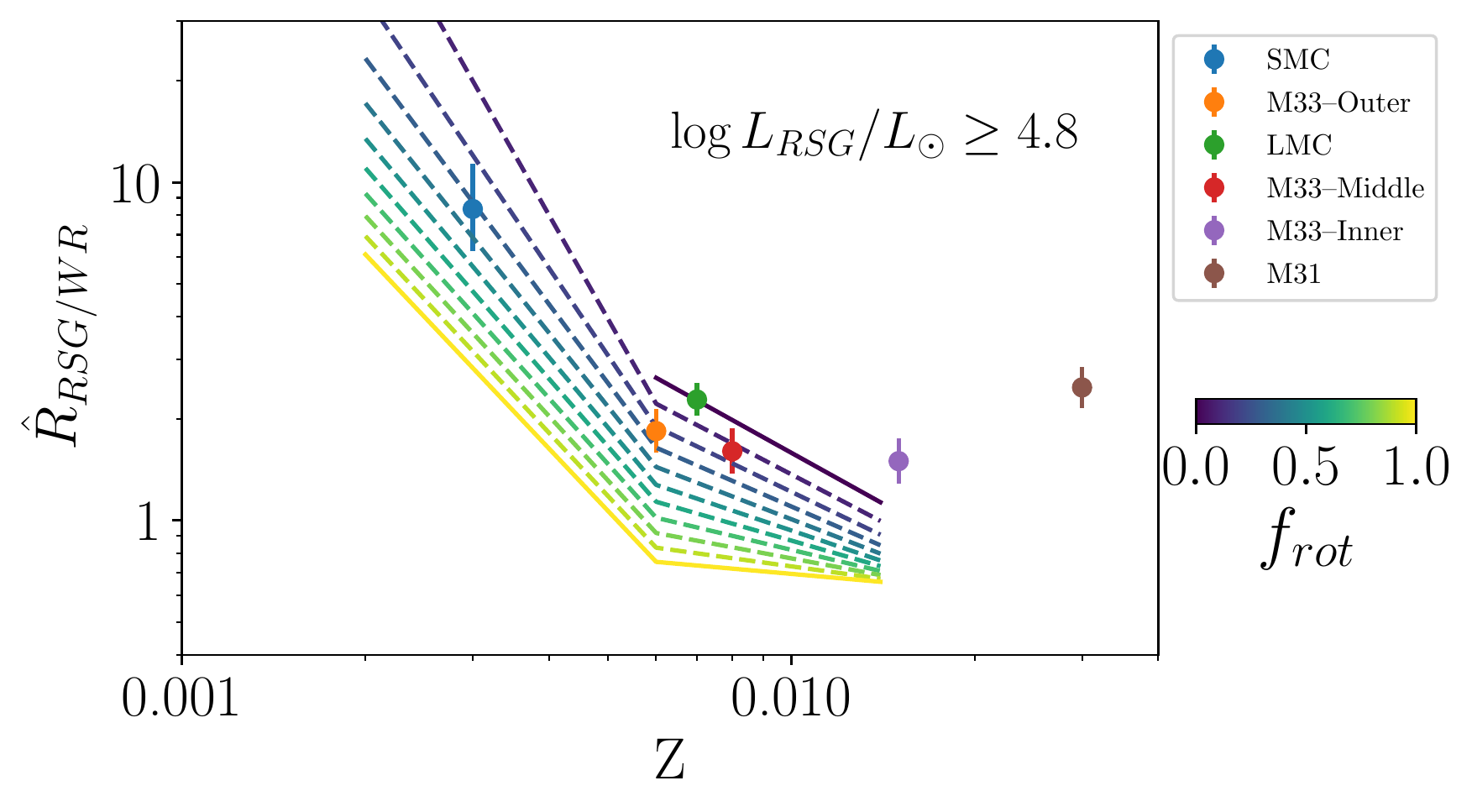}
\includegraphics[width=0.55\textwidth]{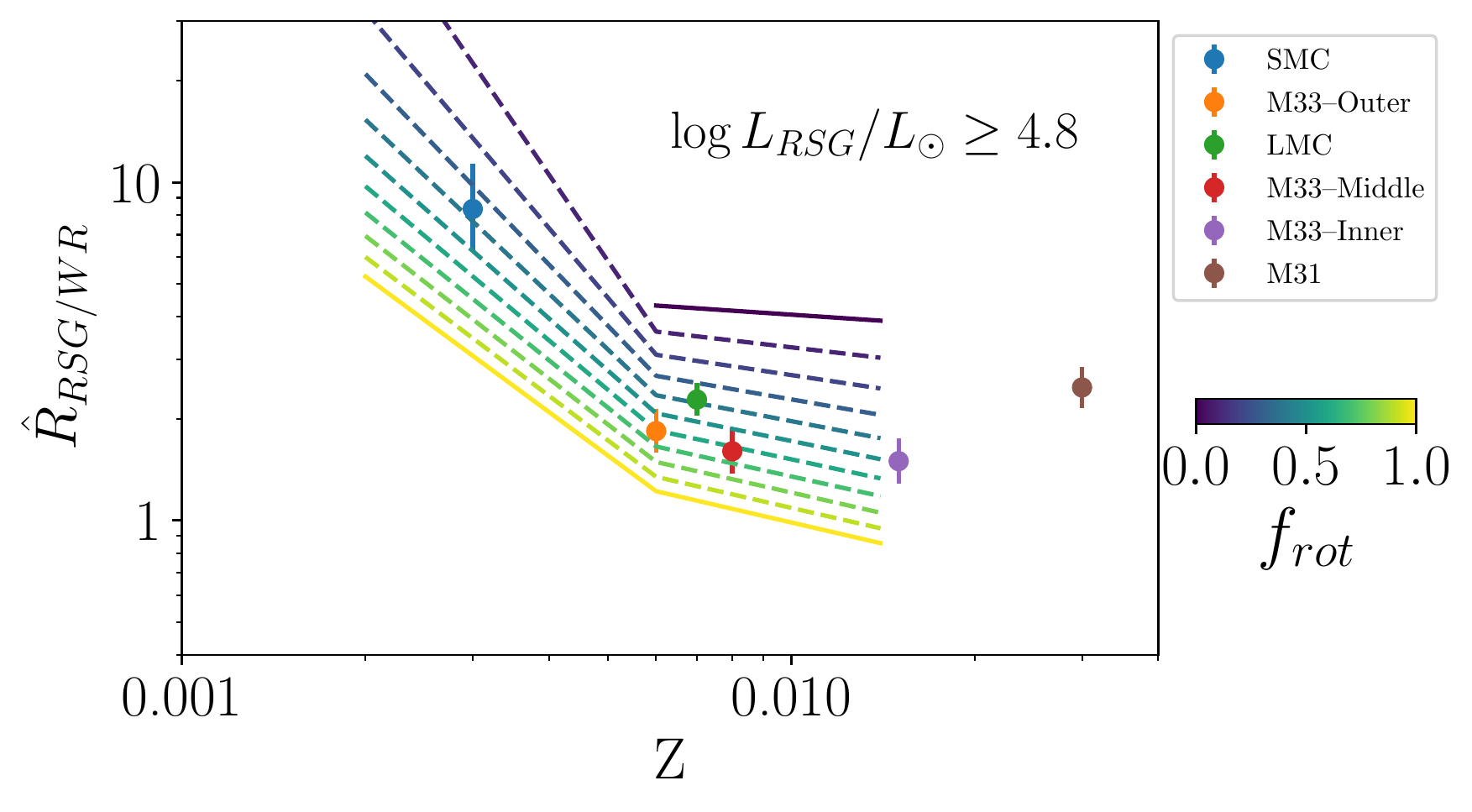}
\caption{\label{fig:Geneva} Comparison with the Geneva single-star evolutionary tracks.  We show the observed RSG/WR ratios (points) compared with the predictions of the Geneva evolutionary tracks (lines).  On the left, we have identified the RSG phases in the models using the RSG effective temperature scale derived from photometry; on the right, we use the RSG effective temperature scale derived from spectroscopic analysis.  The color range of the models correspond to interpolating between the Geneva non-rotating models ($f_{\rm rot}=0$) and the rotating models, and the full-rotating models ($f_{\rm rot}=1$); we have used dashes for the interpolated models.  The latter correspond to an initial rotation velocity of 40\% of the breakup speed.  We note that the effects with respect to rotation are unlikely to be completely linear (see, e.g., Figs.\ 8 and 14 in \citealt{2018ApJS..237...13L}), and the scaling is therefore approximate.
}
\end{figure}

\begin{figure}
\includegraphics[width=0.55\textwidth]{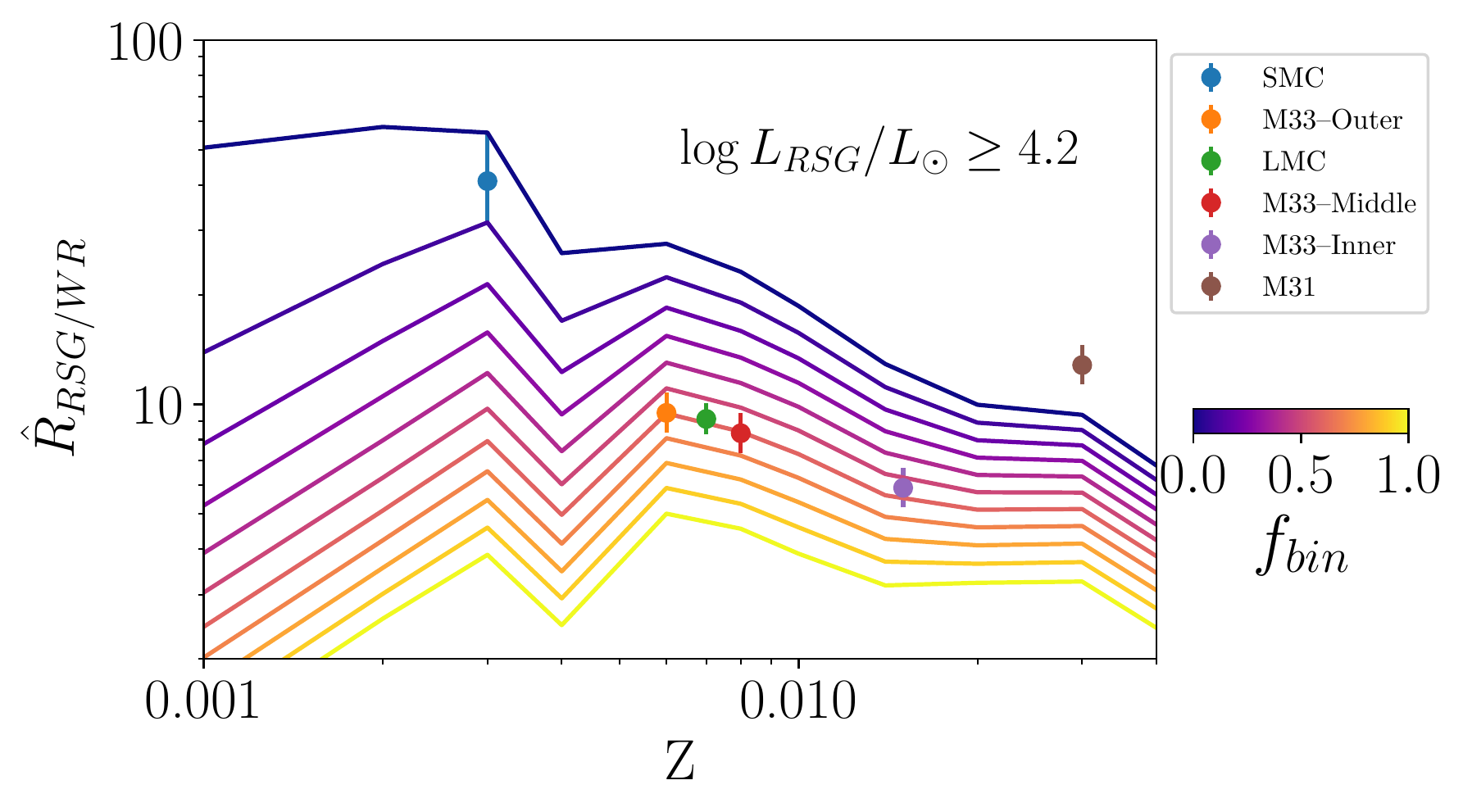}
\includegraphics[width=0.55\textwidth]{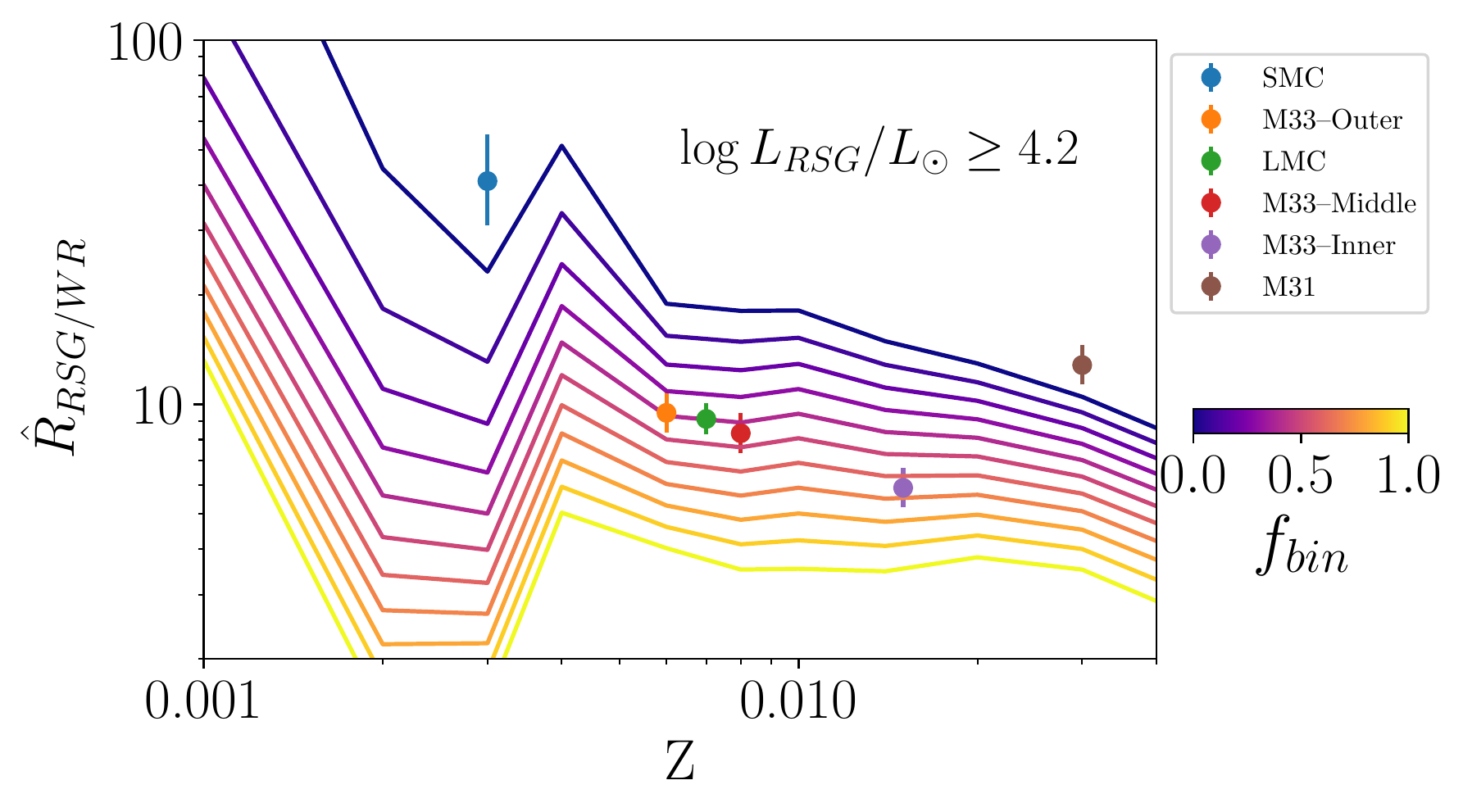}
\includegraphics[width=0.55\textwidth]{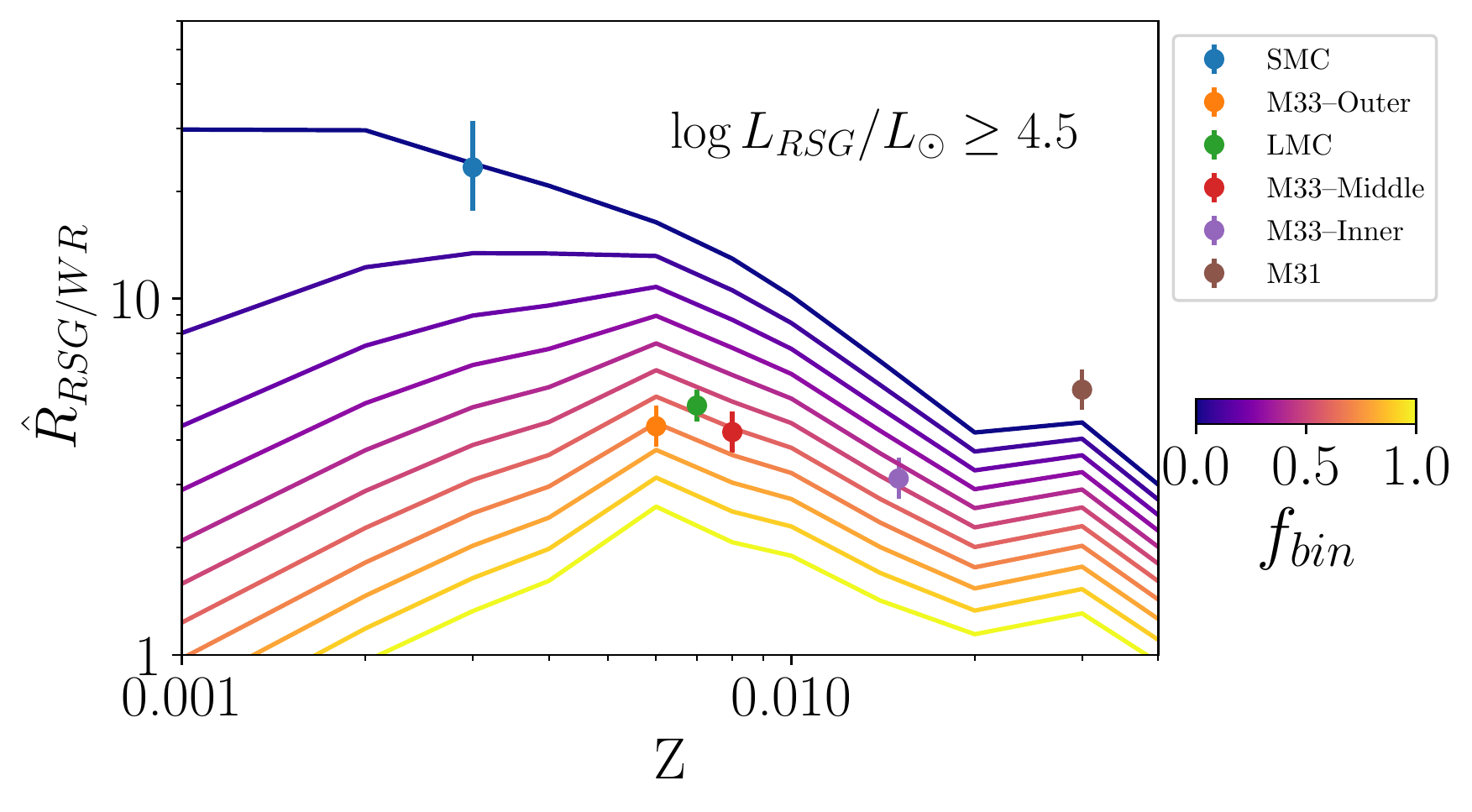}
\includegraphics[width=0.55\textwidth]{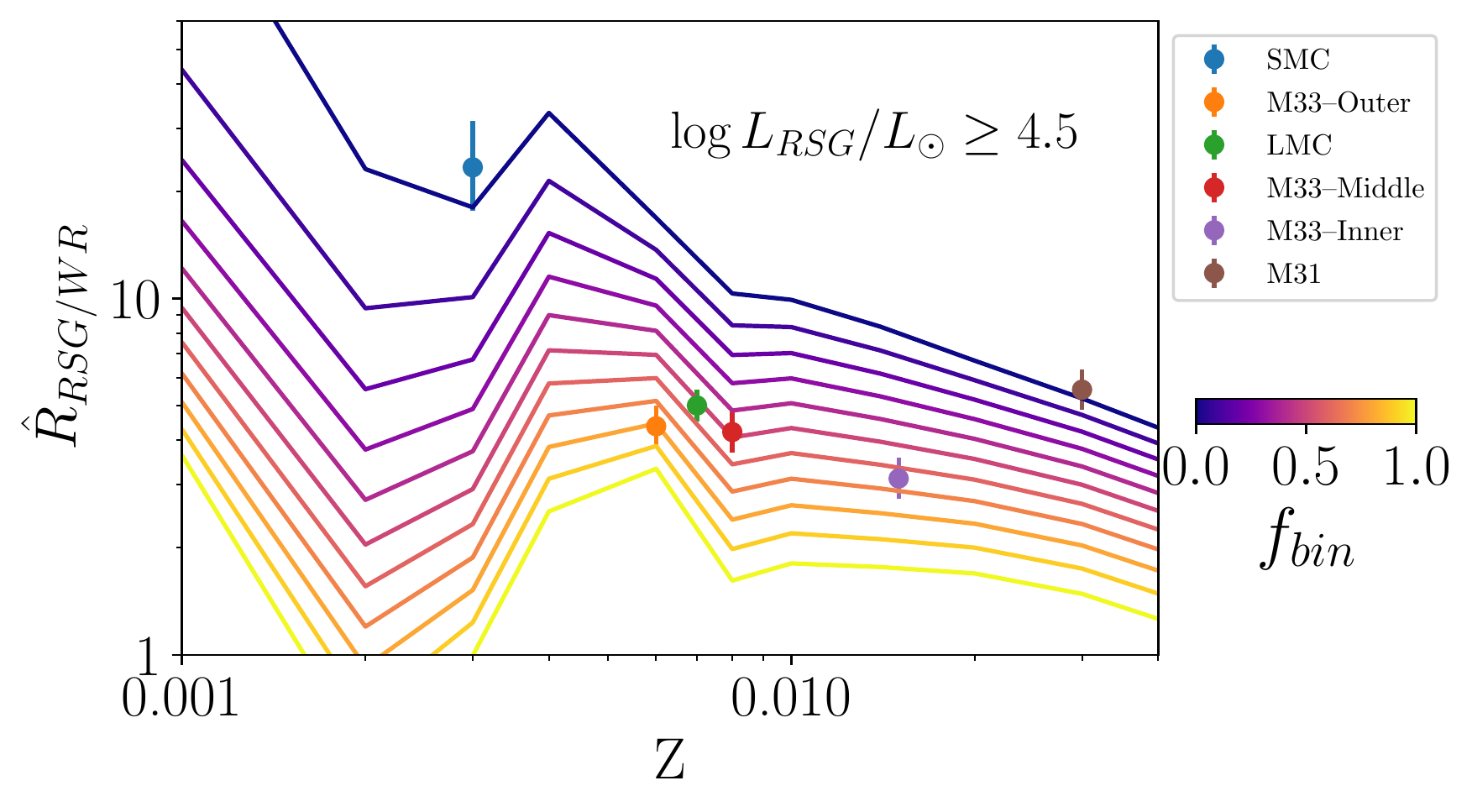}
\includegraphics[width=0.55\textwidth]{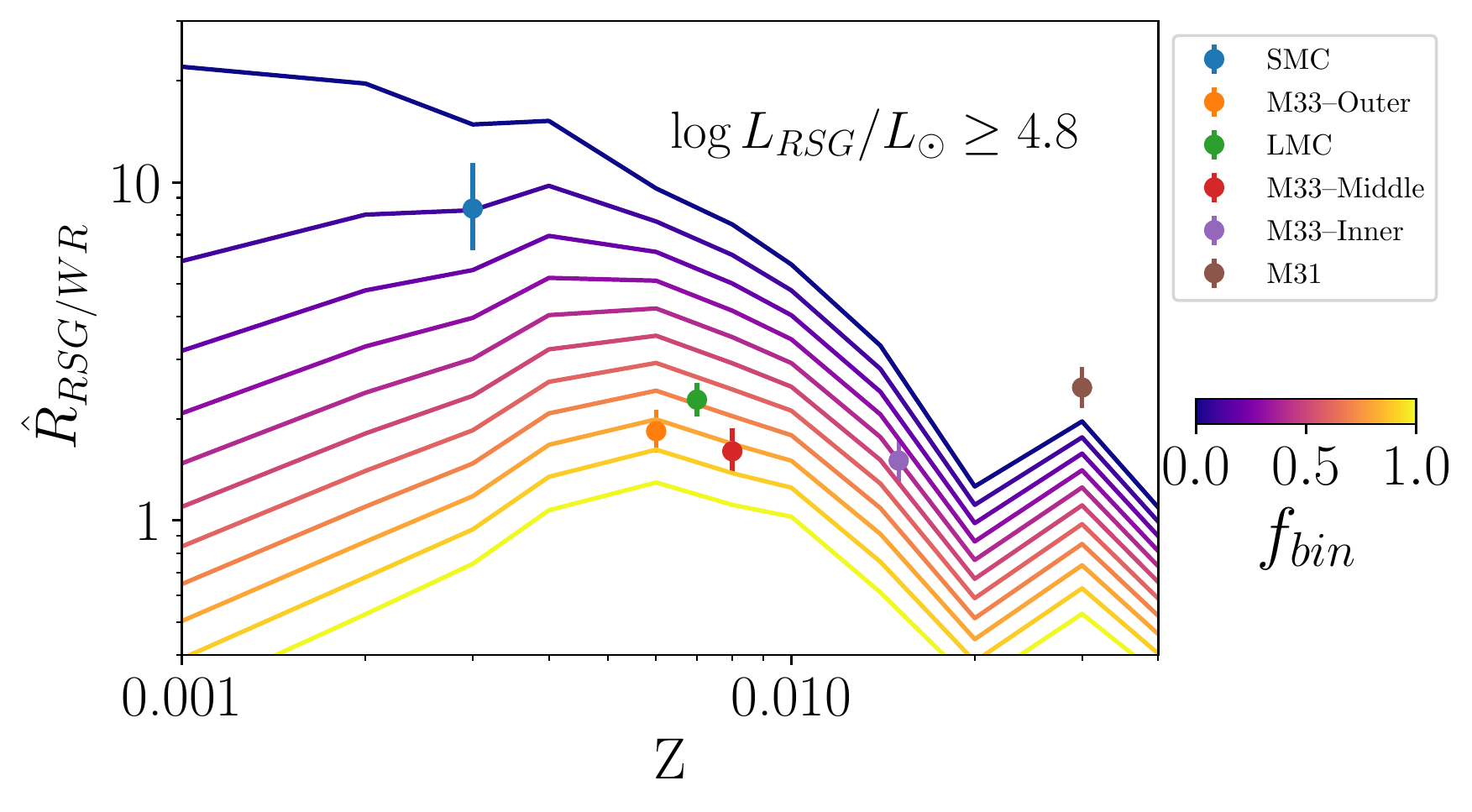}
\includegraphics[width=0.55\textwidth]{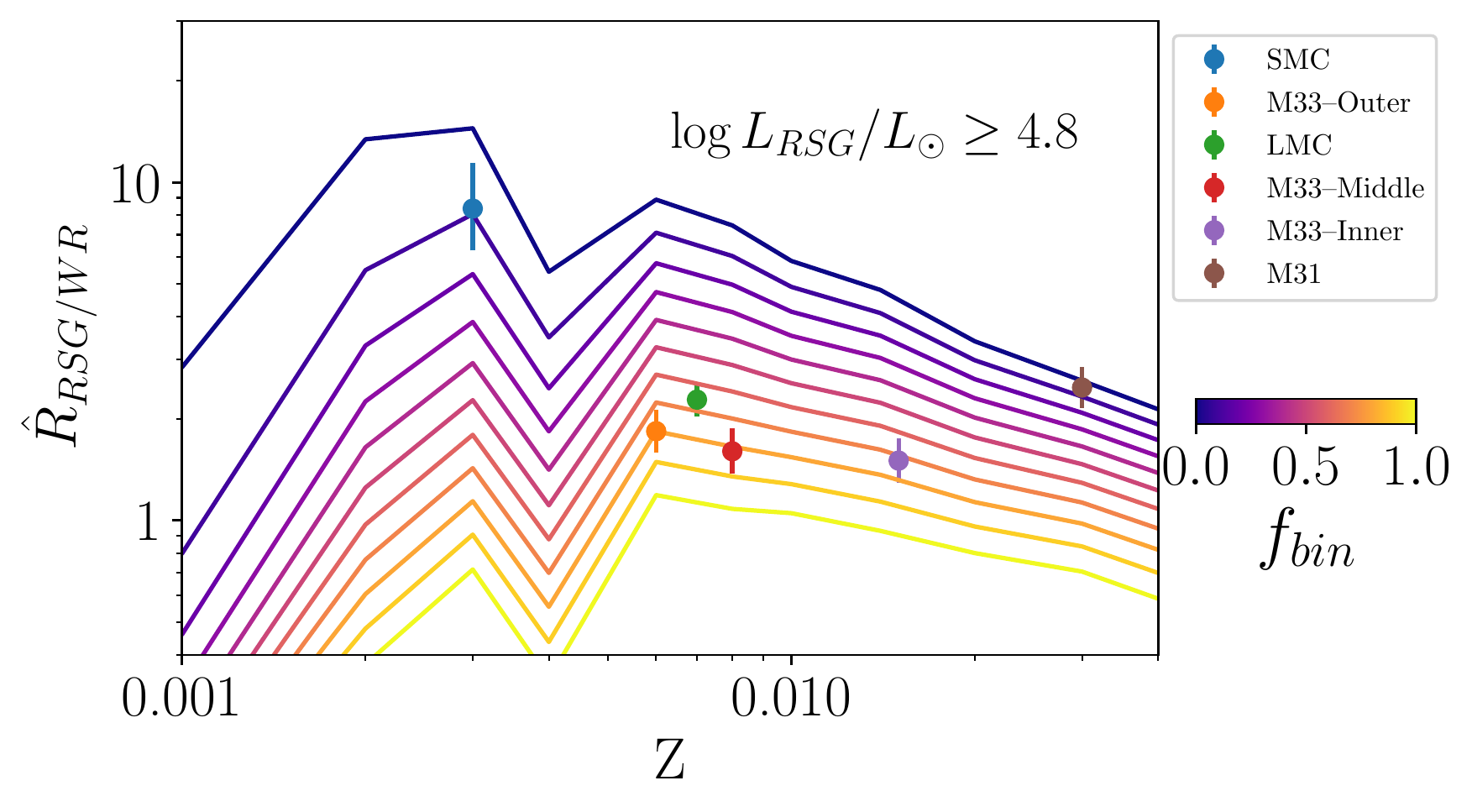}
\caption{\label{fig:BPASS} Comparison with the BPASS evolutionary tracks.  We show the observed RSG/WR ratios (points) compared with the predictions of the BPASS evolutionary tracks (lines).  On the left, we have identified the RSG phases in the models using the RSG effective temperature scale derived from photometry; on the right, we use the RSG effective temperature scale derived from spectroscopic analysis.  The color range of the models correspond to interpolating between the BPASS models with no binaries ($f_{\rm bin}=0$) and the models containing all binaries ($f_{\rm bin}=1$).
}
\end{figure}

\begin{figure}
\includegraphics[width=1.0\textwidth]{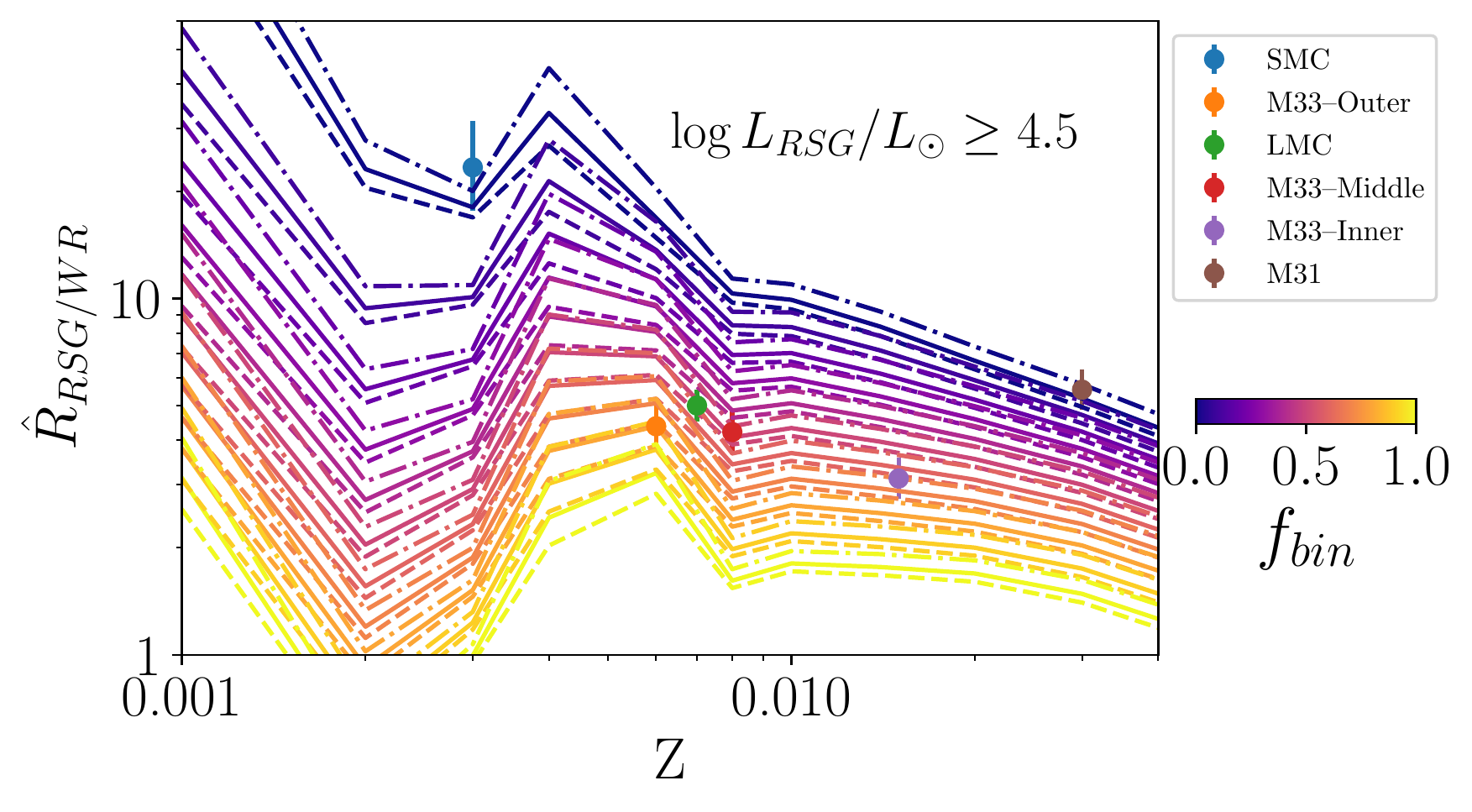}
\caption{\label{fig:20} Effect a steady change in the SFR. We show a comparison of a change of a steady $\pm$30\% change in the SFR over the past 40~Myr on the expected RSG/WR ratio, using BPASS models with a RSG luminosity criteria of $\log L/L_\odot >4.5$ and the temperature limits corresponding to the spectroscopic criteria. The effect is barely discernible, with the upper dashed lines representing a 30\% drop in the SFR to more recent times, and the lower dot-dashed curve showing the effect of a 30\% increase in the SFR to more recent times.}
\end{figure}

\begin{figure}
\includegraphics[width=1.00\textwidth]{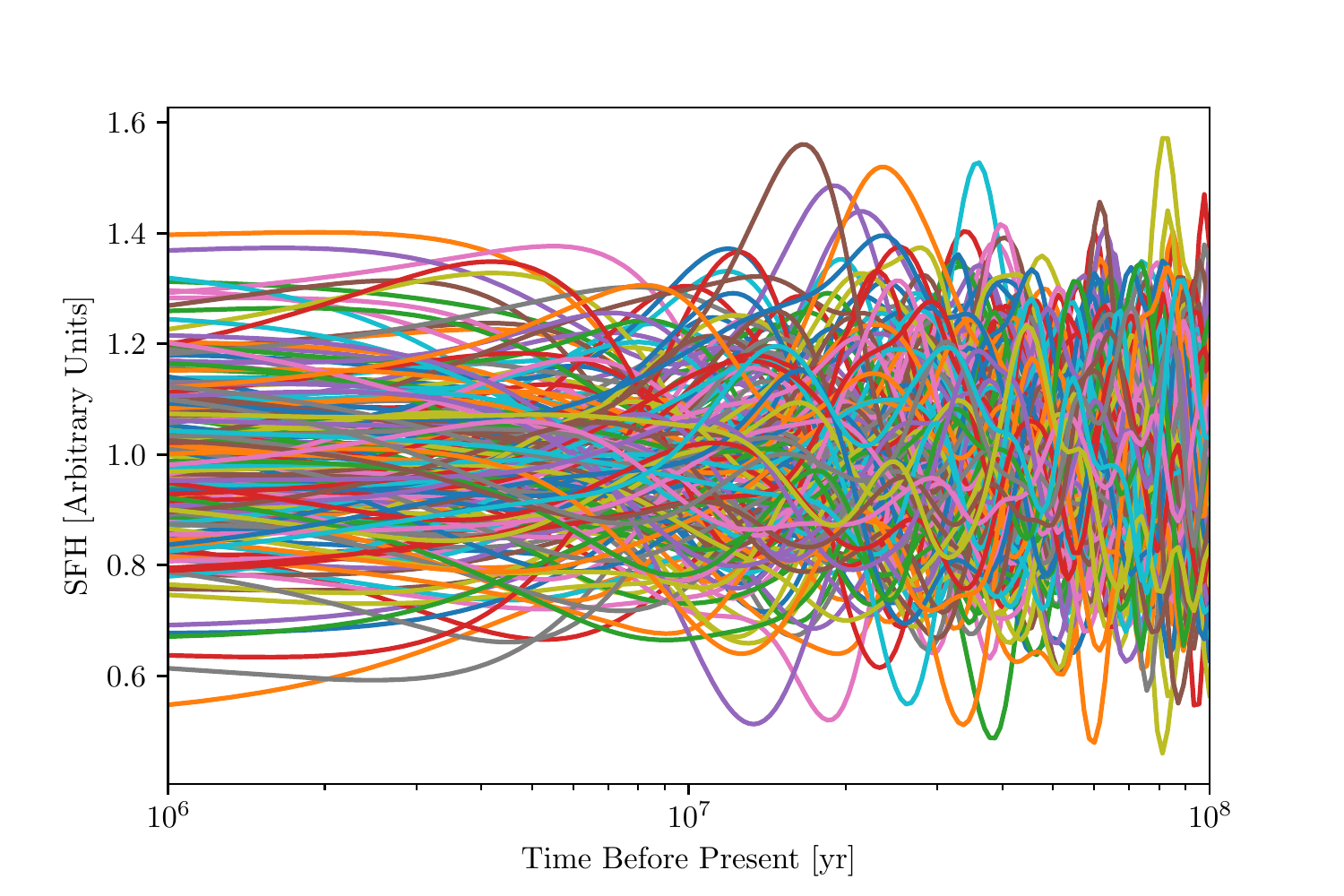}
\includegraphics[width=1.00\textwidth]{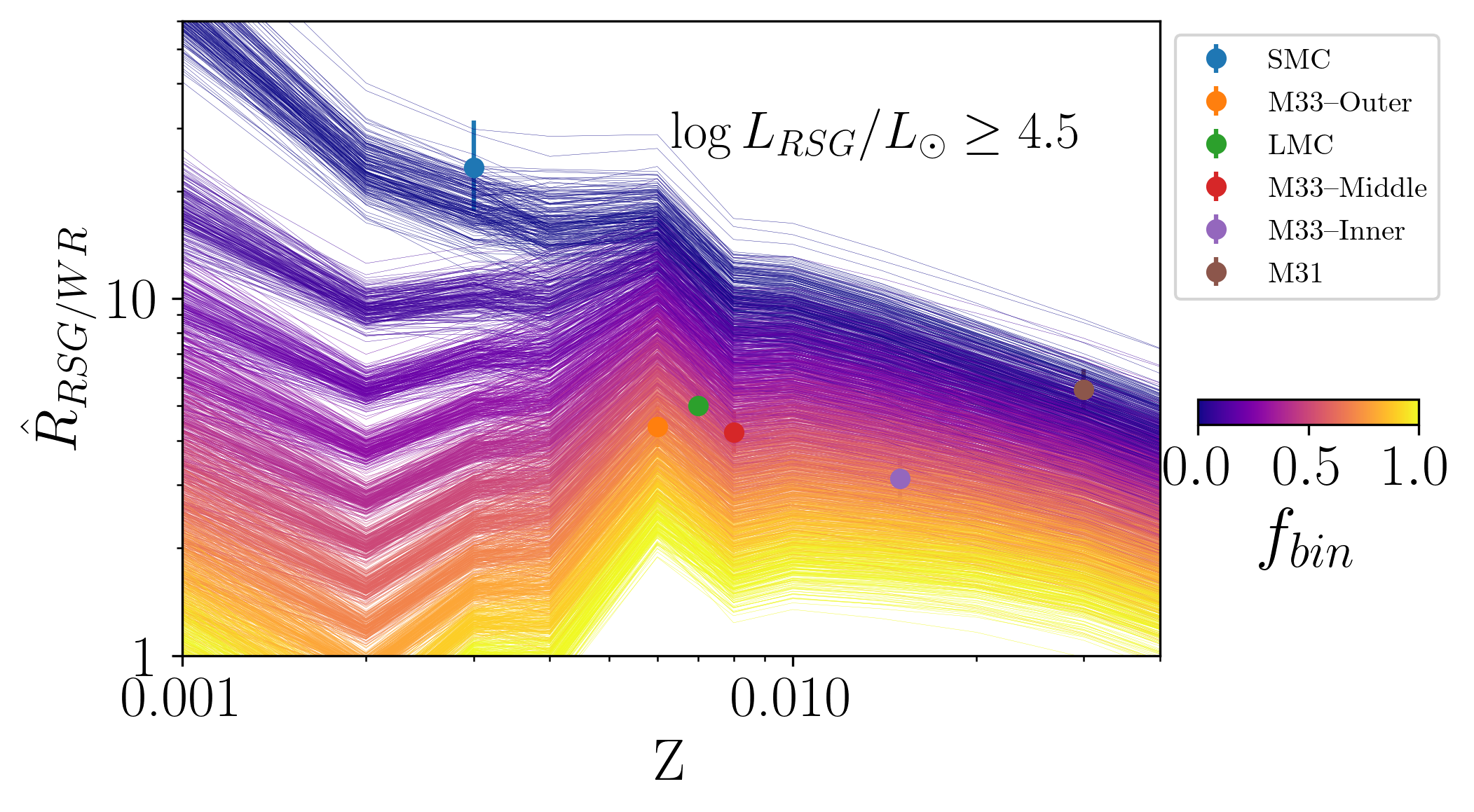}
\caption{\label{fig:yarn} Effect of stochastic changes in the SFR.  The upper figure shows a more realistic model of variations in the SFR, where we have varied the value by 10-40\% on time-scales of 5~Myr.  The lower figure shows the effect on the expected RSG/WR ratios from the BPASS models with a RSG luminosity criteria of $\log L/L_\odot >4.5$ and the temperature limits corresponding to the spectroscopic criteria.}
\end{figure}

In order to identify the RSGs in the SMC, we used the same technique \citet{LMCBins} recently applied to the LMC.  We began by selecting objects from the VizieR version of the 2MASS point-source catalog \citep{2MASS}.  We used a search radius of 3$^{\circ}$ centered on $\alpha_{\rm 2000}$=01:08:00.0 $\delta_{\rm 2000}$=-73:10:00 in order to match the area searched for WRs \citep{FinalCensus}; see Figure~\ref{fig:smc}.   We used only the stars with photometric quality flags of ``AAA," and artifact contamination flags of ``000." We restricted the sample to stars with $K_s\le13$ and $J-K_s\geq0.5$, much fainter and more blue than we expect to find RSGs at the distance of the SMC (59~kpc, \citealt{vandenbergh2000}), as shown below. The resulting CMD is shown in Figure~\ref{fig:SMCCMD}(a).  

As emphasized in earlier papers, one of the concerns is contamination by foreground stars.  We therefore used {\it Gaia} proper motions and parallaxes to separate SMC stars from those in the foreground. The procedure is described in detail in \citet{LMCBins}.  As in that study, we retain stars with ambiguous results or incomplete Gaia data. Owing to the timing of our work, we relied primarily on the Gaia Data Release 2 (DR2, \citealt{DR2}), but as mentioned below, in some cases we also consulted Data Release 3 (DR3, \citealt{DR3}).  Out of 15,005 stars in the photometry list, 9419 (62.8\%) are considered to be certain members, 4903 (32.6\%) are foreground, 415 (2.8\%) have ambiguous results, and 268 (1.8\%) lack Gaia complete Gaia data.  We show the effects of removing the foreground stars in Figure~\ref{fig:SMCCMD}(b). As in the LMC, the vast majority of the contaminants are at the warmer temperatures, outside the region we will use to define the RSGs.  

Of greater concern is the separation of RSGs and AGBs.  As discussed above, \cite{Yang2019} showed that RSGs and AGBs could be readily separated in a ($J-K_s$, $K_s$) CMD, and we have followed that procedure here.  Figure~\ref{fig:WOW} shows
the distribution of stars in such a CMD after foreground stars have been removed, where
we have indicated the region we have adopted for the RSGs, as well as identifying the AGB sequences defined by \citet{2011AJ....142..103B}. We list the photometric criteria adopted in Table~\ref{tab:FunFacts}.

Finally, we need to covert this photometry to the physical properties of effective temperatures and luminosities. As discussed above, we rely upon the {\sc marcs} models \citep{Marcs75,Marcs08,Marcs92} computed with surface gravities appropriate for RSGs ($\log g\sim0.0$) as described in \citet{Levesque2005}.  Following \citet{LMCBins,UKIRT}, we found a simple functional relationship between effective temperatures and the dereddened $(J-K)_0$ colors.  Note that the model colors were computed on the standard \citealt{BessellBrett} system and that we must first transform our 2MASS colors to this system using the relationships described by \citealt{Carpenter}, as given in Table~\ref{tab:FunFacts}.   The previously analyzed RSGs have average $A_V$ values of 0.75 (see \citealt{EmilyMC} and discussion in \citealt{LMCBins,UKIRT}), and we used this value to deredden the transformed colors, using the relationships give by \citet{schlegel}.  The exception were a handful of stars that are too bright to be AGBs, but redder than most of the RSGs; as in previous papers we assume these are RSGs with extra circumstellar reddening \citep{Smoke}, and correct for extinction by following the reddening line back to the mean relationship between $J-K_s$ and $K_s$.  As note in earlier papers, the luminosity we derive is actually quite insensitive to the reddening, as the extinction in $K$ is only 12\% that in $V$.  The relationships and transformations are all summarized in Table~\ref{tab:FunFacts}.   The relationships between our color criteria, and the effective temperatures, have already been given in Table~\ref{tab:RSGcriteria}.  The typical uncertainty in our values are 150~K in $T_{\rm eff}$ and 0.05~dex in $\log L/L_\odot$.
 Of course, as described above, we suspect that the actual ``evolutionary" temperatures are 200~K cooler than these. 

The final source list of SMC RSGs is given in Table~\ref{tab:SourceListSMC}.  There are 1741 RSGs brighter than our $K_s =13.0$ limit.  This corresponds to a completeness limit of $\log L/L_\odot = 3.7$, much less luminous than we are concerned with here. 

In Table~\ref{tab:SourceListSMC} we have included spectral types for a variety of sources.  \citet{EmilyMC} reclassified a number of the RSGs previously identified by \citet{Elias1985} and \citet{MasseyOlsen}, and fit their spectra to refine the effective temperature scale and determine luminosities.  Subsequently, \citet{HV11423} and \citet{EmilyVariables} identified 12 highly unusual RSGs in the Magellanic Clouds, five of which were in the SMC.  These stars underwent large swings in spectral types,  changing from early K-type to M4 or later on the time scale of a few years.  In their cool state, these stars are much later in type than expected for their host galaxies.   In their search for candidate Thorne-\.{Z}ytkow objects, \citet{TZO} examined these and other stars with similarly high $J-K_s$ colors for abundance anomalies, leading to their finding of HV 2112 as the first such candidate.  
When we cross-referenced our list with the recent spectroscopy study of \citet{Dorda2018} we noticed two additional possible spectrum variables. The star 2MASS J01093824-7320024 ([M2002] SMC 069886) was classified as K5-M0~I by \citet{EmilyMC}, but as a very late M-type star (M4 Iab) by \citet{Dorda2018}.  It was subsequently classified as M2~I by \citet{LMCBins}.   The star 2MASS J00485182-7322398 ([M2002] SMC 011939 may be a similarly case, as \citet{EmilyMC} called it a K2~I while \citet{Dorda2018} classified it as M2 Ia-Iab.

The main discrepancy we noted between researchers was that \citet{Dorda2018} classified a number of stars in our list as late G-type.  Four of these had previously been classified by \citet{EmilyMC}, whose classifications range from K1~I to K2-3~I.  
Here we merely note this systematic issue; resolving this is well beyond the scope of the present paper.  We will further note
that one of the stars  (2MASS J00382421-7410196) that was originally in our RSG list was classified by \citet{Dorda2018} as an early M dwarf (M1~V).  Its SMC membership was ambiguous from DR2.  Comparison with the improved astrometry from DR3, however,  confirms it is likely foreground, and we have eliminated it both from our table and from our counts.  

We also included spectral types by \citet{NeugentRSGBinII,LMCBins} as part of their search for RSGs with binary companions, and noted if radial velocity variations were detected by \citet{DordaRV}.  Spectral types of the recently studied stars in NGC 330 by \citet{N330} were also included.  We note that cross-references to names and previous data are not intended to be complete; however, the 2MASS designations can readily be used in SIMBAD to find additional information.

We noticed that very few spectroscopically confirmed RSGs are missing in our list.  2MASS 01000932-7208441 ([M2002] SMC 48122) was classified by \citet{EmilyMC} as K1 I,
but its 2MASS photometry is flagged as unreliable in $J$ and $K_s$.  Although the RSG+B binary 2MASS J00464984-7313525 ([M2002] SMC 7618) was originally identified as probable foreground from the Gaia DR2 astrometry, DR3 fixed the problem as the parallax went from $0.99\pm0.25$ to $0.13\pm0.08$, and we have retained the star in our source list.  Finally, 9 of the 16 NGC 330 RSGs classified by \citet{N330} are missing, as crowding in the extreme center of the cluster compromised the 2MASS photometry.   By contrast, there are 260 stars in Table~\ref{tab:SourceListSMC}.  Thus the requirement that we accept only the best 2MASS photometry in identifying RSGs in the SMC has had a very minimal effect on the statistics. 

\clearpage

\clearpage

\begin{figure}
\includegraphics[width=0.48\textwidth]{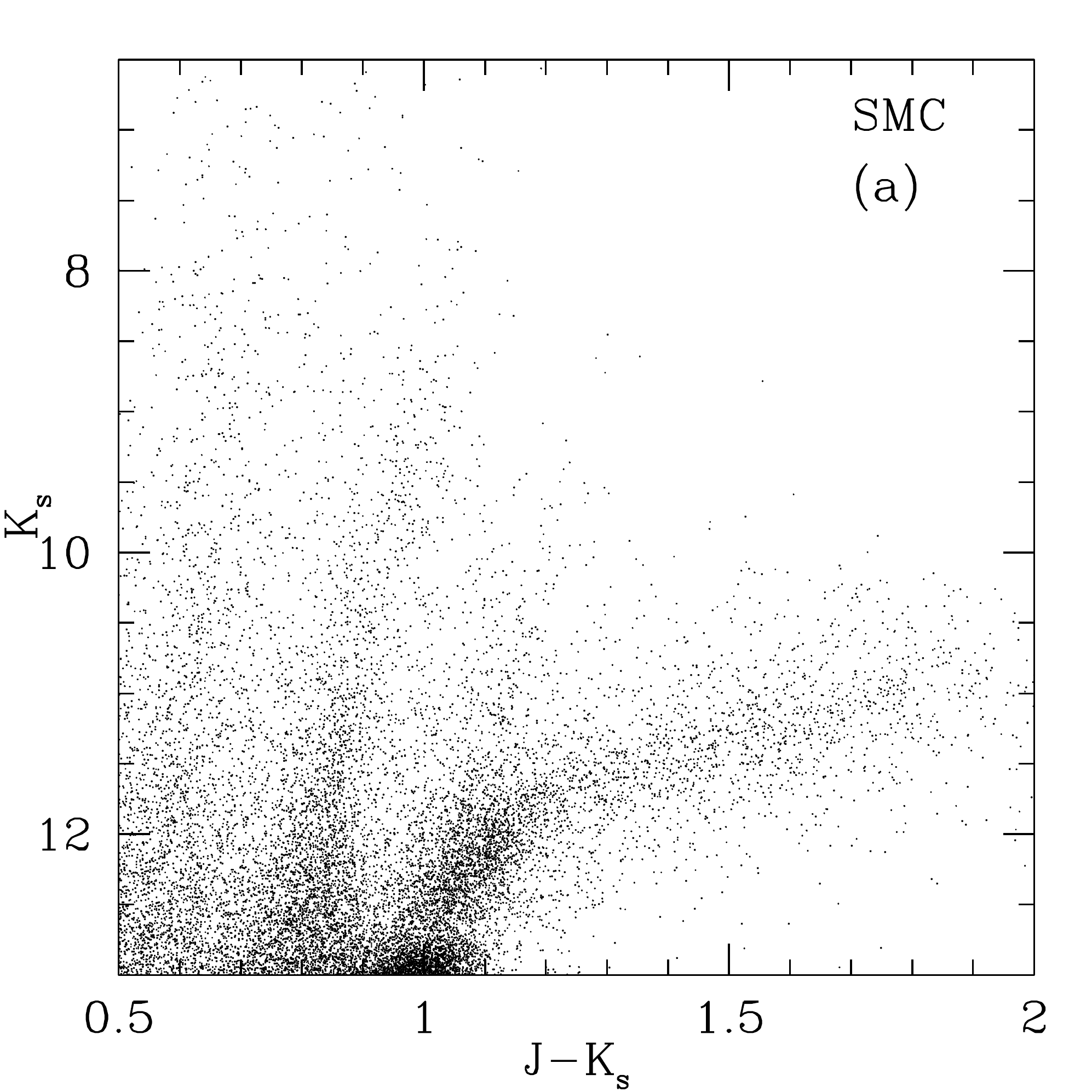}
\includegraphics[width=0.48\textwidth]{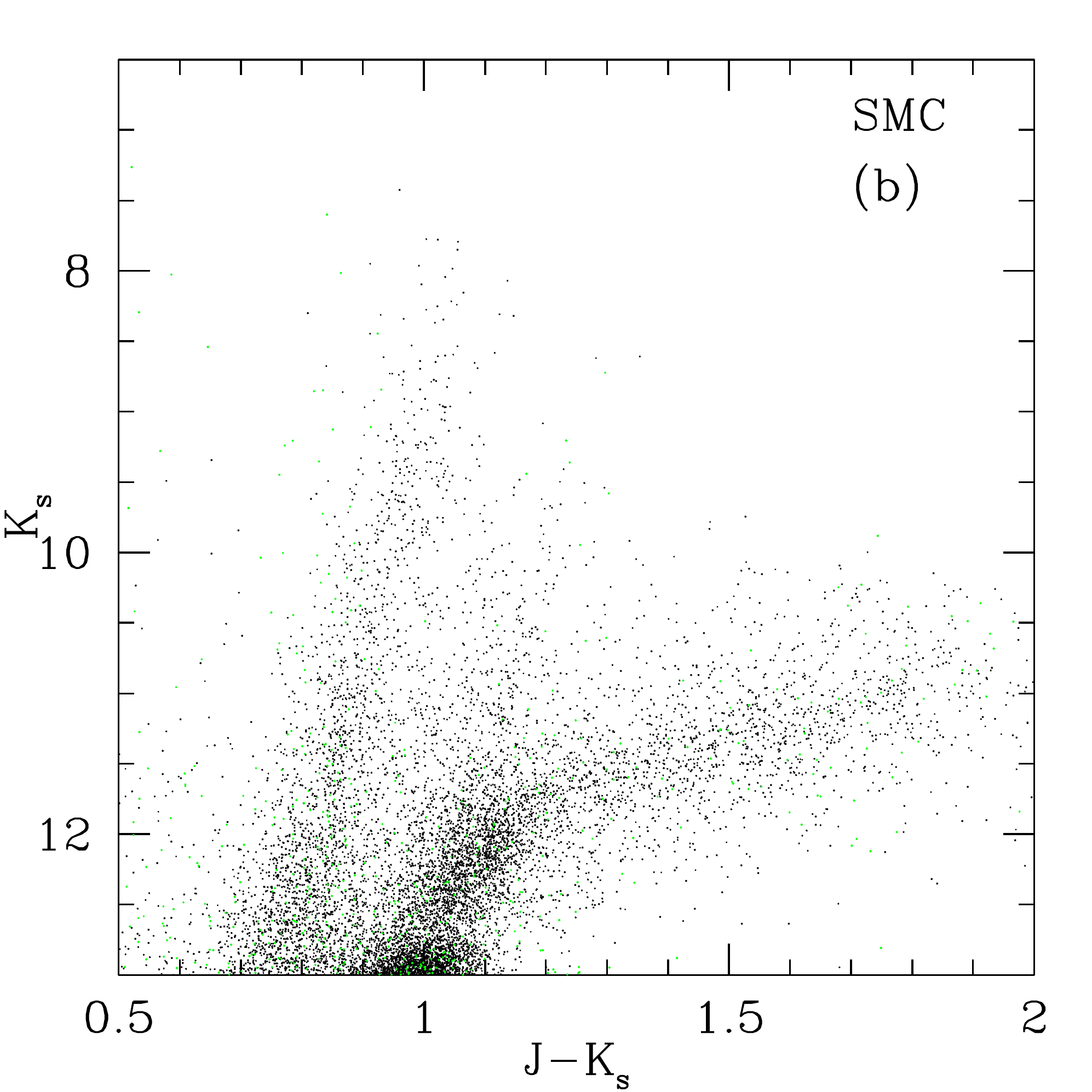}
\caption{\label{fig:SMCCMD} The CMD for our SMC sample.  (a) The CMD is shown for all 15,005 stars in the initial 2MASS sample. (b) The same as (a) but now with probable foreground stars removed.  Green points denote the stars without complete (or any) Gaia data.}
\end{figure}

\begin{figure}
\includegraphics[width=1.0\textwidth]{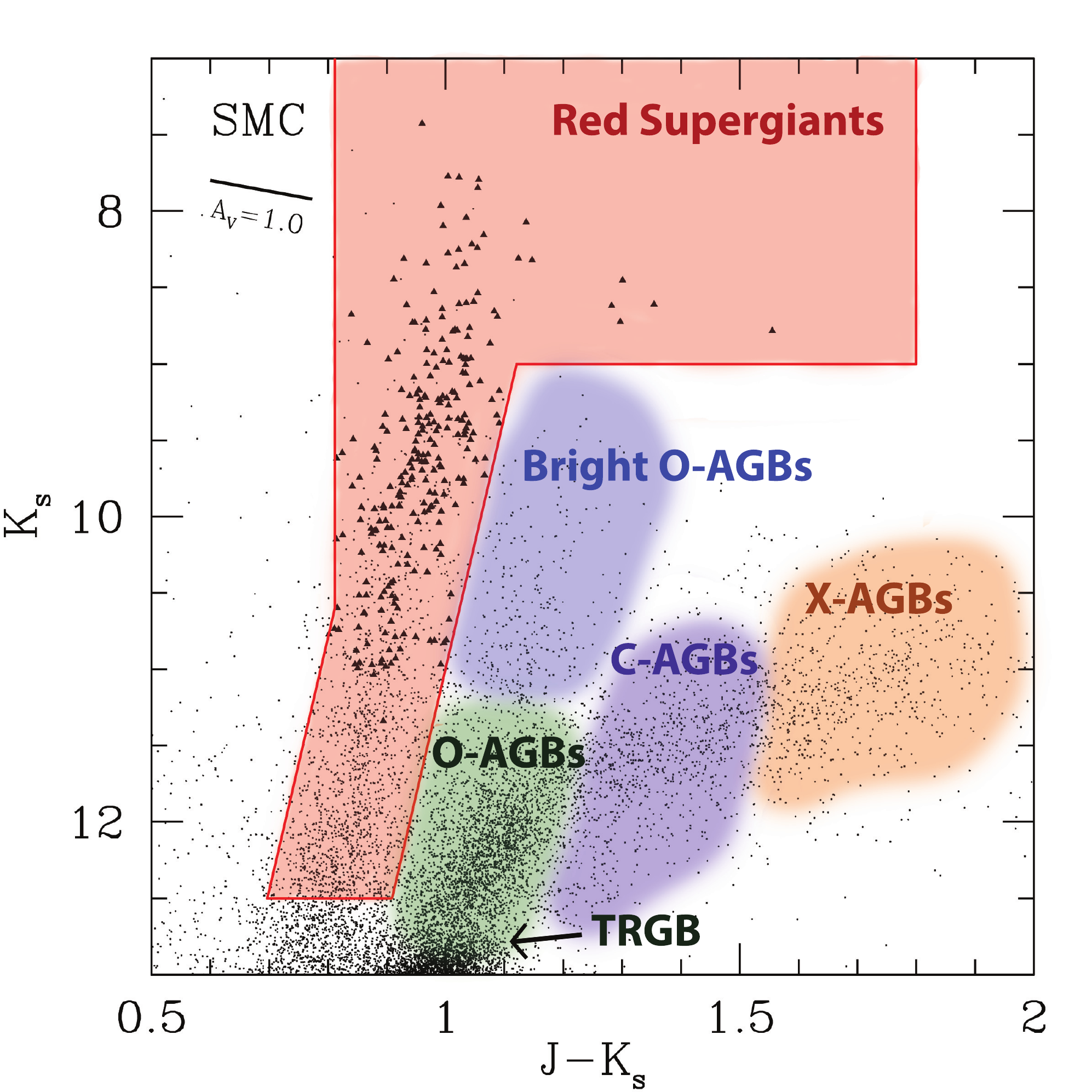}
\caption{\label{fig:WOW} The CMD for cool members of the SMC.  The various AGB branches from \citet{2011AJ....142..103B} are labeled,  along with the tip of the red giant branch (TRGB).  The triangles show the red supergiants that have some spectral information in Table~\ref{tab:SourceListSMC}. The reddening vector corresponding to $A_V=1.0$ is shown. This figure may be compared to a similar one for the LMC in \citet{LMCBins} (their Figure 15).}
\end{figure}

\begin{deluxetable}{l l l}
\tablecaption{\label{tab:RSGcriteria} Temperature Selection of RSGs}
\tablehead{
\colhead{Galaxy}
&\colhead{Relations}
&\colhead{Ref.}
}
\startdata
SMC:& & 1\\
       &For $\log L/L_\odot  < 4.32,  T_{\rm phot} < 5613 - 279.1 \log L/L_\odot$ \\
       &For $\log L/L_\odot \geq 4.32,  T_{\rm phot} < 4407$\\
       & For $\log L/L_\odot < 4.80,   T_{\rm phot} > 5223-279.4 \log L/L_\odot$\\
       & For $\log L/L_\odot \geq4.80,   T_{\rm phot} > 3880$\\
LMC+outer M33: & & 2,3\\
       & For $\log L/L_\odot < 4.28, T_{\rm phot}<5736-360.3 \log L/L_\odot$\\
       &For $\log L/L_\odot \geq 4.28, T_{\rm phot}< 4195$ \\
       &For $\log L/L_\odot < 4.78, T_{\rm phot}>5326-360.0 \log L/L_\odot$\\
       &For $\log L/L_\odot \geq4.78, T_{\rm phot}>3604$\\
 M33--inner+middle: & & 3\\
 	&For $\log L/L_\odot < 4.77, T_{\rm phot} < 5831-359.7 \log L/L_\odot$\\
	&For $\log L/L_\odot \geq4.77, T_{\rm phot} < 4114$\\
	&For $\log L/L_\odot < 4.78, T_{\rm phot} > 5324-359.7 \log L/L_\odot$\\
	&For $\log L/L_\odot \geq 4.78, T_{\rm phot} > 3604$\\	
M31:  & & 3\\
	&For $\log L/L_\odot < 4.47, T_{\rm phot} < 5767-388.7 \log L/L_\odot$\\
	&For $\log L/L_\odot \geq 4.47, T_{\rm phot} < 4030$\\
	&For $\log L/L_\odot < 4.73, T_{\rm phot} > 5249-388.7 \log L/L_\odot$\\
	&For $\log L/L_\odot \geq 4.73, T_{\rm phot} > 3412$\\
For all galaxies: & & 1,2\\
	&$T_{\rm spect}=T_{\rm phot}-200$ \\
\enddata
\tablerefs{1--This paper; 2--\citealt{LMCBins}; 3--\citealt{M3133RSGs}
\tablecomments{$T_{\rm phot}$ refers to the effective temperature determined from the de-reddened $J-K_s$ photometry.
$T_{\rm spect}$ refers to the effective temperature determined from spectroscopic analyses using the TiO band. See Section~\ref{Sec-physical} of text.}
}
\end{deluxetable}
\begin{deluxetable}{l c c c c}
\tablecaption{\label{tab:Numbers} Number of RSGs and WRs}
\tablewidth{0pt}
\tablehead{
\colhead{Galaxy}
&\multicolumn{3}{c}{\#RSGs}
&\colhead{\#WRs}   \\  \cline{2-4}
&\colhead{$L/L_\odot \geq 4.2$}
&\colhead{$L/L_\odot \geq 4.5$}
&\colhead{$L/L_\odot \geq 4.8$} \\
&\colhead{(9$M_\odot$)}
&\colhead{(12$M_\odot$)}
&\colhead{(15$M_\odot$)} \\
}
\startdata
SMC          & \phn501 &  283 &    101 &      \phn12 \\
LMC(total)   & 1233 & 711 &    316 &     142 \\
\multicolumn{1}{l}{\phm{XX}No 30Dor/III\tablenotemark{a}} 
&1014   &556    &253   &   111 \\
M31\tablenotemark{b}           &1854   &758    &357   &    148 \\
\multicolumn{1}{l}{\phm{XX}Low Ext.\ 50\%} & 885 & 383 & 171 & 69 \\
M33   &1648   &816     &344   &    209 \\
\multicolumn{1}{l}{\phm{XX}Inner\tablenotemark{c}}    
& \phn425   &226     &108    &   \phn72 \\
\multicolumn{1}{l}{\phm{XX}Middle\tablenotemark{d}}&  
\phn567 &  288 &    109&      \phn68 \\
\multicolumn{1}{l}{\phm{XX}Outer\tablenotemark{e}}&  
\phn656  & 302  &   127 &      \phn69 \\
\enddata
\tablenotetext{a}{After removing stars
within 10\arcmin\ of 
$\alpha_{\rm 2000}$=05:38:42.40, $\delta_{\rm 2000}$=-69:06:03.4 (30~Dor),
and within 45\arcmin\ of 
$\alpha_{\rm 2000}$=05:31:24.33, $\delta_{\rm 2000}$=-66:54:45.0 (Constellation~III).}
\tablenotetext{b}{$\rho\leq0.75$}
\tablenotetext{c}{Inner: $\rho<0.25$}
\tablenotetext{d}{Middle: $0.25\leq \rho < 0.5$}
\tablenotetext{e}{Outer: $0.5 \leq \rho \leq 1.0$}
\end{deluxetable}

\begin{deluxetable}{l c c c c c c}
\tablecaption{\label{tab:Ratios} Number Ratios as a Function of Metallicity}
\tablewidth{0pt}
\tablehead{
\colhead{Galaxy}
&\colhead{$\log$(O/H)+12}  
&\colhead{$Z$}
&\multicolumn{3}{c}{$\hat{R}_{\rm RSGs/WRs}$}  \\ \cline{4-6}
&&&\colhead{$L/L_\odot \geq 4.2$}
&\colhead{$L/L_\odot \geq 4.5$}
&\colhead{$L/L_\odot \geq 4.8$} \\
&&&\colhead{(9$M_\odot$)}
&\colhead{(12$M_\odot$)}
&\colhead{(15$M_\odot$)}
}
\startdata
SMC & 8.00 & 0.003 & $41.0^{+14.0}_{-10.0}$ & $23.3^{+8.2}_{-5.7}$ & $8.4^{+3.0}_{-2.1}$ \\ 
M33--Outer\tablenotemark{a} & 8.29 & 0.006 & $9.5^{+1.3}_{-1.1}$ & $4.4^{+0.6}_{-0.5}$ & $1.8^{+0.3}_{-0.3}$ \\ 
LMC(total) & 8.38 & 0.007 & $8.7^{+0.8}_{-0.7}$ & $5.0^{+0.5}_{-0.4}$ & $2.2^{+0.2}_{-0.2}$ \\ 
\multicolumn{1}{l}{\phm{XX}No 30Dor/III\tablenotemark{b}} & 8.38 & 0.007 & $9.1^{+1.0}_{-0.9}$ & $5.0^{+0.6}_{-0.5}$ & $2.3^{+0.3}_{-0.2}$ \\ 
M33--Middle\tablenotemark{c} & 8.41 & 0.008 & $8.3^{+1.2}_{-1.0}$ & $4.2^{+0.6}_{-0.5}$ & $1.6^{+0.3}_{-0.2}$ \\ 
M33--Inner\tablenotemark{d} & 8.72 & 0.015 & $5.9^{+0.8}_{-0.7}$ & $3.1^{+0.5}_{-0.4}$ & $1.5^{+0.2}_{-0.2}$ \\ 
M31\tablenotemark{e} & 8.90 & 0.030 & $12.5^{+1.1}_{-1.0}$ & $5.1^{+0.5}_{-0.4}$ & $2.4^{+0.2}_{-0.2}$ \\ 
\multicolumn{1}{l}{\phm{XX}Low Ext.\ 50\%} & 8.90 & 0.030 & $12.8^{+1.8}_{-1.5}$ & $5.5^{+0.8}_{-0.7}$ & $2.5^{+0.4}_{-0.3}$ \\ 
\enddata
\tablenotetext{a}{M33--Outer: $0.5 \leq \rho \leq 1.0$}
\tablenotetext{b}{After removing stars
within 10\arcmin\ of2
$\alpha_{\rm 2000}$=05:38:42.40, $\delta_{\rm 2000}$=-69:06:03.4 (30~Dor),
and within 45\arcmin\ of 
$\alpha_{\rm 2000}$=05:31:24.33, $\delta_{\rm 2000}$=-66:54:45.0 (Constellation~III).}
\tablenotetext{c}{M33--Middle: $0.25\leq \rho < 0.5$}
\tablenotetext{d}{M33--Inner: $\rho<0.25$}
\tablenotetext{e}{$\rho\leq0.75$}
\end{deluxetable}

\clearpage
\begin{deluxetable}{l l r}
\tabletypesize{\scriptsize}
\tablecaption{\label{tab:FunFacts} Adopted and Derived Relations for the SMC}
\tablewidth{0pt}
\tablehead{
&\colhead{Relation}
&\colhead{Source}
}
\startdata
\sidehead{Adopted Distance:}
\phantom{MakeSomeSpace}&59~kpc (DM=18.9~mag)& 1 \\
\sidehead{Reddening Relations:}
&$A_K =0.12 A_V = 0.686 E(J-K)$ & 2 \\
&$E(J-K) = A_V/5.79 $ & 2 \\
\sidehead{RSG Photometric Criteria:}
        &$10.6<K_s \leq 13.5$: $K_s \geq K_{s0}$ and $K_s \leq K_{s1}$ & 3 \\
        &$K_s\leq10.6$: $J-K_s\geq 0.812$ and $K_s \leq K_{s1}$ & 3 \\
                &$K_s\leq 9.0$ and $(J-K_s)\leq 1.8$: $J-K_s\geq 0.812$ & 3 \\
        &$K_{s0}=24.00-16.50(J-K_s)$  & 3,4\\
        &$K_{s1}=27.50-16.50(J-K_s)$  & 3,4\\
\sidehead{Adopted Extinction:}
        & $A_V=0.75$ & 3 \\
\sidehead{Conversion of 2MASS ($J, K_s$) to Standard System ($J, K$):}
&$K=K_s + 0.044$ & 5\\
&$J-K = (J-K_s+0.011)/0.972$ & 5\\
\sidehead{Conversion to Physical Properties (Valid for 3500-4500 K):}
& $T_{\rm eff} = 5592.5- 1656.0 (J-K)_0$ & 3\\
&${\rm BC}_K = 5.495 - 0.73697 \times T_{\rm eff}/1000$ & 3\\
&$K_* = K - A_K$ & \nodata \\
&$M_{\rm bol} = K_* + {\rm BC}_K - DM$ & 1 \\
&$\log L/L_\odot =(M_{\rm bol}-4.75)/-2.5$ & \nodata \\
\enddata
\tablerefs{1--\citealt{vandenbergh2000}; 2--\citealt{Schlegel1998};
3--This paper; 4--\citealt{2006AA...448...77C}; 5--\citealt{Carpenter}
}
\end{deluxetable}

\clearpage

\begin{deluxetable}{c c c r r r r c c c l r l l}
\rotate
\tabletypesize{\scriptsize}
\tablecaption{\label{tab:SourceListSMC} RSG Source List for the SMC}
\tablewidth{0pt}
\tablehead{
&&&&&&&&&&&\multicolumn{2}{c}{Spectral Type} \\  \cline {11-12}
\colhead{2MASS}
&\colhead{$\alpha_{\rm J2000}$}  
&\colhead{$\delta_{\rm J2000}$}
&\colhead{$K_s$}
&\colhead{$\sigma_{Ks}$}
&\colhead{$J-K_s$}
&\colhead{$\sigma_{J-Ks}$}
&\colhead{Gaia\tablenotemark{a}}
&\colhead{$T_{\rm eff}$\tablenotemark{b,c}}
&\colhead{$\log L/L_\odot$\tablenotemark{b,d}}
&\colhead{Type} &\colhead{Ref.}
&\colhead{Other ID}
&\colhead{Comments} 
}
\startdata
J00450138-7350513 &00 45 01.383 &-73 50 51.34 &11.170 &0.019 &0.918 &0.032 &0&4200 &4.06 &\nodata &\nodata&\nodata&\nodata\\
J00450313-7255156 &00 45 03.135 &-72 55 15.61 &10.314 &0.023 &0.935 & 0.033 &0&4200 &4.39& G8 Ib &     5 &[GDN2015] SMC079&\nodata\\
J00450456-7305276 &00 45 04.566 &-73 05 27.69  &8.071 &0.023 &1.137 &0.033 &0& 3850 &5.19 &M2 I & 1& [M2002] SMC 005092 &RV var Ref 9\\
J00450482-7340132 &00 45 04.824 &-73 40 13.28 &12.209 &0.027 &0.791 &0.036 &0&4450 &3.71&\nodata&\nodata&\nodata&\nodata\\
J00450746-7327417 &00 45 07.464 &-73 27 41.79 &10.300 &0.025 &1.008 &0.035 &0&4050 &4.36&\nodata&\nodata&\nodata&\nodata\\
J00450758-7358102 &00 45 07.581& -73 58 10.21 &10.868 &0.025 &0.923 &0.035 &0&4200 &4.18 &G7 Iab-Ib & 5 &[GDN2015] SMC080&\nodata\\
J00450816-7325382 &00 45 08.165 &-73 25 38.22 &12.131 &0.029 &0.849 &0.038 &0&4350 &3.71&\nodata&\nodata&\nodata&\nodata\\
J00450831-7254146 &00 45 08.312 &-72 54 14.64 &12.273 &0.032 &0.920 &0.042 &0&4200 &3.62&\nodata&\nodata&\nodata&\nodata\\
\enddata
\tablecomments{Table~\ref{tab:SourceListSMC} is published in its entirety in the machine-readable format.  A portion is shown here for guidance regarding its form and content.}
\tablenotetext{a}{Membership flag based on Gaia information: 0=Probable member; 1=Uncertain member; 2=Incomplete Gaia data.}
\tablenotetext{b}{Computed assuming Av=0.75 mag.}
\tablenotetext{c}{Typical uncertainties 150~K.}
\tablenotetext{d}{Typical uncertainties 0.05~dex.}
\tablerefs{1--\citet{EmilyMC}; 
2--\citet{HV11423}; 
3--\citet{EmilyVariables}; 
4--\citet{TZO}; 
5--\citet{Dorda2018}; 
6--This paper and references therein; 
7--\citet{LMCBins}; 
8--\citet{NeugentRSGBinII}; 
9--\citet{DordaRV}; 
10--\citet{N330};
11--\citet{1980ApJ...242L..13E};
12--\citet{2018MNRAS.479.3101B}.
}
\end{deluxetable}
\clearpage

\bibliographystyle{aasjournal}
\bibliography{masterbib.bib}

\end{document}